\documentclass[10pt]{article}
\usepackage{graphicx}
\usepackage{epsfig}
\usepackage{color}

\begin{document}

\title{The building up of individual  inflexibility in opinion dynamics}
\author{Andr\'e C. R. Martins\\
GRIFE -- EACH -- Universidade de S\~ao Paulo,\\
Rua Arlindo B\'etio, 1000, 03828--000,  S\~ao Paulo, Brazil\\
\vspace{0.35cm}\\
Serge Galam\\
Centre de Recherche en \'Epist\'emologie Appliqu\'ee,\\
\'Ecole Polytechnique and CNRS, \\CREA, Blvd Victor, 32,
75015 Paris, France}

\date{(amartins@usp.br, serge.galam@polytechnique.edu)}

\maketitle

\begin{abstract}
Two models of opinion dynamics are entangled in order to build a more realistic model of inflexibility. The first one is the Galam Unifying Frame (GUF), which incorporates rational and inflexible agents, and the other one considers the combination of Continuous Opinions and Discrete Actions (CODA). While initially in GUF, inflexibility is a fixed given feature of an agent, it is now the result of an accumulation for a given agent who makes the same choice through repeated updates. Inflexibility thus emerges as an internal property of agents becoming a continuous function of the strength of its opinion. Therefore an agent can be more or less inflexible and can shift from inflexibility along one choice to inflexibility along the opposite choice. These individual dynamics of the building up and falling off of an agent inflexibility are driven by the successive local updates of the associated individual opinions. New results are obtained and discussed in terms of predicting outcomes of public debates.

\end{abstract}


\section{Introduction}

Most models of opinion dynamics use local rules of opinion updates to monitor the dynamics of individual agents \cite{castellanoetal07}. The agents are often identical and can shift opinion as a function of the chosen update rule. Indeed many models are based on the sole existence of an update rule which is legitimated by the psycho-sociological interpretation of the rule. For instance the Sznajd model evokes the postulated principle  "United we Stand, Divided we Fall"  \cite{sznajd00,stauffer03a} while  Galam models consider the democratic principle "one person one vote"  \cite{galam04b}. However such an approach can be misleading precisely with respect to the psycho-sociological consequences of selecting a model. It has been shown that all discrete models, i.e., different claimed psycho-sociological mechanisms, are indeed identical and can be mapped to the Galam Unifying Frame (GUF) \cite{galam05b}. They can also be obtained as limit cases \cite{martins12a} of a generalized Continuous Opinions and Discrete Actions (CODA) model \cite{martins08a,martins08b} .
 
More recently a good deal of works has been devoted to the incorporation of different types of agents considering heterogeneous agents \cite{galam05}. For instance, the effect of contrarian agents who always oppose the majority choice, either local or global was studied \cite{galam04,schneider04a,delalamaetal05a,zhongetal05a,borghesigalam06a,kurten08a,jacobsgalam08a,martinskuba09a}. A small proportion of these contrarian agents were shown to drive a total upside down of the dynamics from a threshold dynamics onto a threshold-less dynamics with a unique attractor located at exactly fifty-fifty. This feature was used to explain the so called hung elections, a new phenomenon repeatedly observed in the last ten years in democratic societies \cite{galam04}. 

In addition, the effect of inflexible agents who never shift opinion have been also investigated. The idea is to account for the fact that in the real social world some agents are observed to be strongly opinionated individuals who do not change opinion whatever opposite argument is given to them. Indeed, they can play a very important role in a public debates. By simply holding to their opinions when meeting different people in various discussing groups, they can cause a minority to slowly increase in size until it becomes a majority and eventually wins the debate. This effect was shown in models of opinion dynamics using inflexible agents  \cite{galamjacobs07}. Inflexibility is found to produce a series of unexpected and disturbing results. Depending on the respective proportions of inflexibles at each side of a debate, even a very small minority could  convince the whole population about its opinion \cite{galam06b}.

Such a feature of agents holding to their opinion against exchanges with others holding the opposite opinion has been introduced first twenty years ago using local individual fields in the modeling of group decision making within the frame of magnetic like models \cite{galammoscovici91}. Opinion dynamics since, has been the subject of numerous works where some sort of inflebility or extremism was introduced and its effects studied \cite{castellanoetal07,deffuantetal00,hegselmannkrause02,weisbuchetal05,galam05b,deffuant06,galam06b,galamjacobs07,martins08a,martins08b,martins10a}.

Nevertheless, it is worth to emphasize that when heterogeneous agents have been incorporated in a model their respective proportions were kept fixed during the dynamics process of emergence of a collective opinion. Accordingly the composition of the system in terms of the various types of agents is given as an external parameter. It is part of the initial condition in the problem. Once the debate is turned on, no rational individuals would become inflexibles, nor inflexibles have a chance of eventually getting convinced they were wrong and in turn shift opinion as it is sometimes observed in the  reality. We denote rationals the corresponding agents who do hold an opinion but are susceptible to shift to another one while discussing with others provided enough convincing arguments are given. 

In this paper we  extend the notion of inflexibility by incorporating the CODA formalism \cite{martins08a,martins08b,martinspereira08a,martins08e,martins12a}, where discrete choices are associated with a strength of the associated opinion. That way, an inflexible can be defined as just someone who has a very strong opinion, which could be the result of many interactions with agents holding the same opinion. It thus becomes difficult to have this agent to shift its own opinion.  Accordingly, inflexibility becomes the result of an internal dynamics of the system which is associated to a given individual. Simultaneously the reverse process holds true. An inflexible agent can have its stubbornness more and more weakened making it a rational agent. Also an inflexible agent can progressively move from inflexibility along one opinion along inflexibility along the opposite opinion. Yet, total inflexibility can be introduced with no difficulty, requiring just that their opinion require a large multiple of the number of allowed interactions for each individual. That way, the dynamics will stop before the inflexibles would actually change their points of view.

\section{Model}

Inflexibles were originally introduced as individuals who were so sure of themselves that their opinion would not change, regardless of social influence \cite{galam05,galamjacobs07}. It makes sense, however, that, given enough time and social pressure, even an inflexible could, in principle, become a floater, with a softer view and capable of being convinced. On the other hand, people who were not so convinced at first could have their own opinions strengthened to the point of being considered inflexibles.  

The changing of status between floaters and inflexibles can be introduced if we use the Continuous Opinions and Discrete Actions (CODA) model \cite{martins08a,martins08b} as basis. In the CODA model, an underlying and non-observable continuous opinion was used to measure how certain each agent was about its decision on which of two options it preferred. That means an agent can have an inner opinion that makes it easy for its choice to change and be a floater. Or the agent could have a very strong opinion, requiring many interactions before any change could be observed in its current opinion, making the agent basically an inflexible. Therefore, we will use the CODA model in order to explore in more depth the possible consequences of the existence of inflexibles among the initial debaters in a public issue. This will extend the previous results by Galam \cite{galamjacobs07} to include the possibility that inflexibles can eventually become floater and vice-versa. 

In the model, $N$  agents will be left to interact freely with each other, with no network structure assumed. Each agent $i$ has a strength of opinion $\nu_i$, such that the spin choice associated with $\nu_i$ will be given by $\sigma_i=sign(\nu_i)$. In the original CODA model, when one agent observed in instant $t$ a neighbor with choice $\sigma_j=\pm 1$, the agent opinion was updated by
\[
\nu_i(t+1) = \nu_i(t)+\sigma_j(t).
\]
This could mean a change of the observed opinion $\sigma_i$, whenever $\nu_i$ changed sign.

Following GUF treatment,  one group of $s$ agents are randomly drawn at each time step, forming a debate group. Unlike the treatments by Galam, the whole population is not divided at each interation step, but only one group is formed. That way, some agents will, in a given simulation, interact more with their peers than others. That is probably closer to how society conducts real debates. Time can be measured as the average number of debates an agent has participated in. 

In the original model from Galam, everyone who is not an inflexible, adopts the opinion of the observed majority in the group, when the group interacts.
Applying the CODA formalism to GUF interactions is simple. Inside a group of size $s$, the agent will observe $s-1$ different agents and we can use a CODA for each other member of the group. While a reasonable proposal, this raises an important question, when comparing with GUF. In GUF, it was the majority that drove the dynamics. And in any majority, the opinion of the agent itself counts. Take, per example, a group of size $s=4$. When there is a tie, what each agent observes is another agent who agrees with him and two other who disagree. By only observing others, there is a majority against its opinion. This will only be a tie if the agent also uses its own opinion in the updating process. In this paper, we will consider that ties have no influence when they happen, with no bias towards one of the options introduced, i.e., at a tie agents keep their respective current opinions.

However, it is not clear whether the agent should actually take its own opinion in the balance for an eventual update. Its own opinion is not a new information, so, in principle, it should only matter as prior knowledge, not as new data. But majority rule model do include the opinion of the agent since the agent does count for the majority. Also, psychological literature suggests that people do change their mind slowly when facing new information\cite{philipsedwards66,kahanetal11}, suggesting it might be a good idea to have a model when the own opinion of the agent is counted for the updating. Since this issue is possibly open to debate, we will study both options here, with and without self influence. That is, when a group of size s is formed, an agent $i$ will update $\nu_i$ according to
\begin{equation}\label{eq:codaguf}
\nu_i(t+1) = \nu_i(t)+p-m,
\end{equation}
where $p$ (number of pluses) is the number of agents in the group with $\sigma_j=+ 1$ and $m$ (number of minuses) is the number of agents with $\sigma_j= -1$. Depending on the case, the agent opinion will be counted towards $p$ and $m$ or not.

We should also notice that in discrete models, the outcome of any meetings of the group is that everyone who is not an inflexible is convinced by the group. Here, however, agents are only inflexibles in the sense that they may have a strong opinion. They change their minds when an update from Equation (\ref{eq:codaguf}) changes the sign of $\nu_i$, which will be harder for some agents and easier for others. That introduces a new question. When a group meets and agreement is not reached between its members, it is natural, under some circumstances, that the debate will go on a little longer. Therefore, we need to introduce another parameter $M$, to measure the maximum number of interactions (debates) each group will agree to keep discussing before they agree to disagree. Obviously, if there is no limit and $M$ is effectively infinite, all debates end in agreement (unless real inflexibles are introduced) regardless of strength of opinions and we recover the discrete case. This can be seen as another example of how traditional discrete models can be understood as limit cases of Bayesian inspired models \cite{martins08e,martins12a}. Here, we will only explore cases where $M$ is fixed and equal for all agents. That means that for the same group being updated, M successive updates are performed before putting back the agents in the whole population, unless consensus is reached first.

\section{Results}

Following the GUF treatment of the problem, the system was implemented by randomly drawing groups of size $g$ from the population of size $n$. However, here, it is not the whole population that is divided, just groups randomly drawn, allowing some agents to interact more often than others by chance. In order to measure time, we use the average number of groups each agent will participate, that is $T$ is increased of one each time $N$ groups divided by the group size have interacted. Once a group is randomly drawn, the opinions of each member change according to the dynamics explained in the previous Section. Both the case where the self opinion counts and doesn't count were implemented.

In the GUF treatment, with discrete opinions, a group of inflexibles were able to change the opinion of the majority when the population had to choose between two competing ideas, $A$ or $B$ (corresponding to $\sigma=+1$ and $\sigma=-1$ respectively). By checking what happens under a different model, not only plan to learn about the models, but also investigate the robustness of the result that the existence of inflexibles can change a majority of initial supporters for one of the ideas. Each simulation is performed with initial conditions such that there is a probability $p_A$ that each agent will support $A$. Normal agents start with $|\nu |=0.5$ so that one interaction can convince them to change their options. However, if an agent supports $A$ (or $B$) , there is a probability $I_A$ (and a possibly different $I_B$) that it is a weak version of an inflexible, that is, its initial opinion will be reinforced by a large amount $E$. If $E$ is an integer, it corresponds to the number of interactions with disagreeing majorities an agent must have before changing its choice.

\subsection{Size 3 groups}

For the simulation results shown here, unless otherwise mentioned, we have taken $E=50$. That makes the agents reasonably inflexible, but not as much as to completely prevent them from changing their decisions over the course of many interactions.

\begin{figure}
\centering
\begin{tabular}{ccc}
\epsfig{file=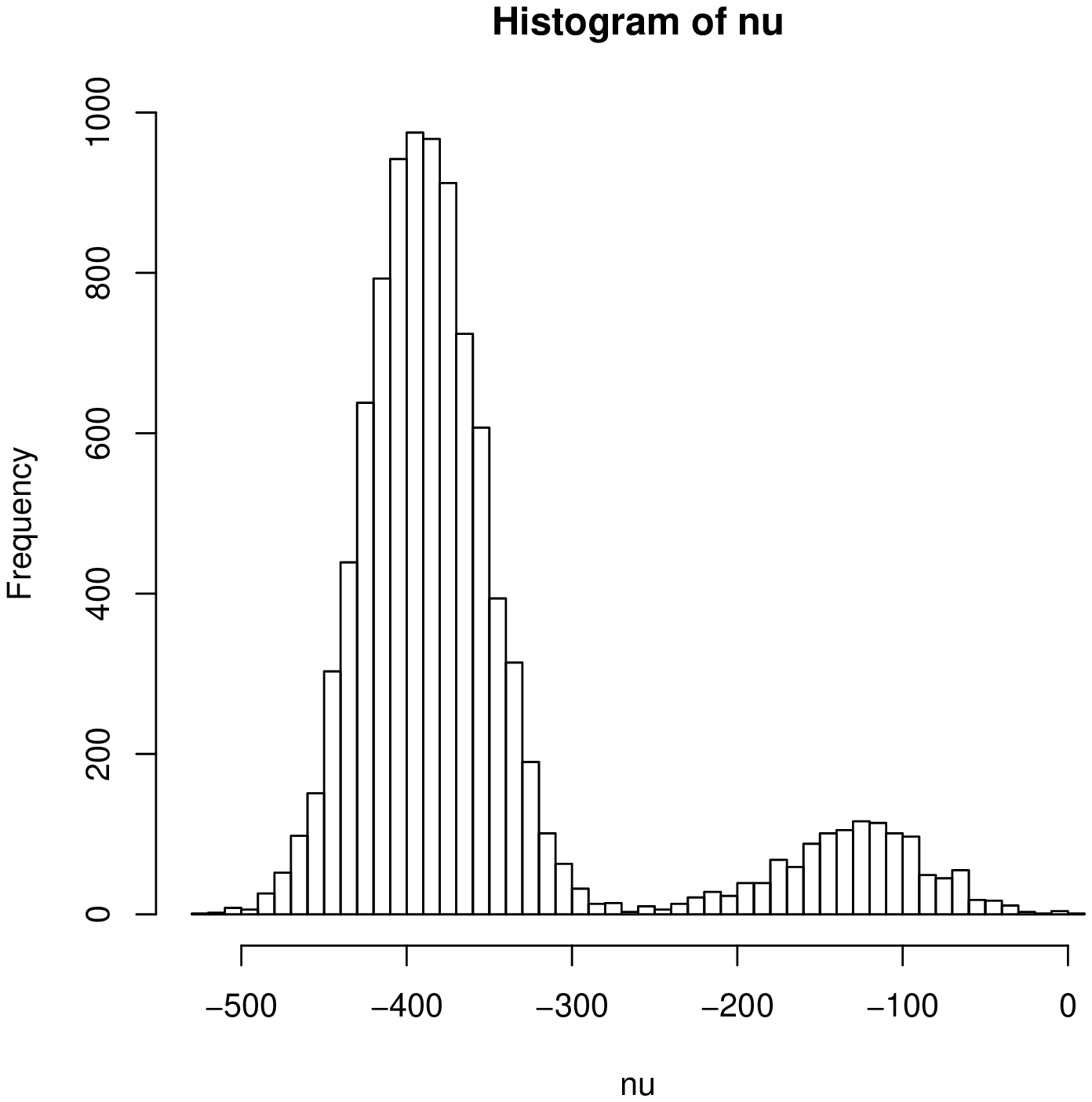,width=0.3\linewidth,clip=} & 
\epsfig{file=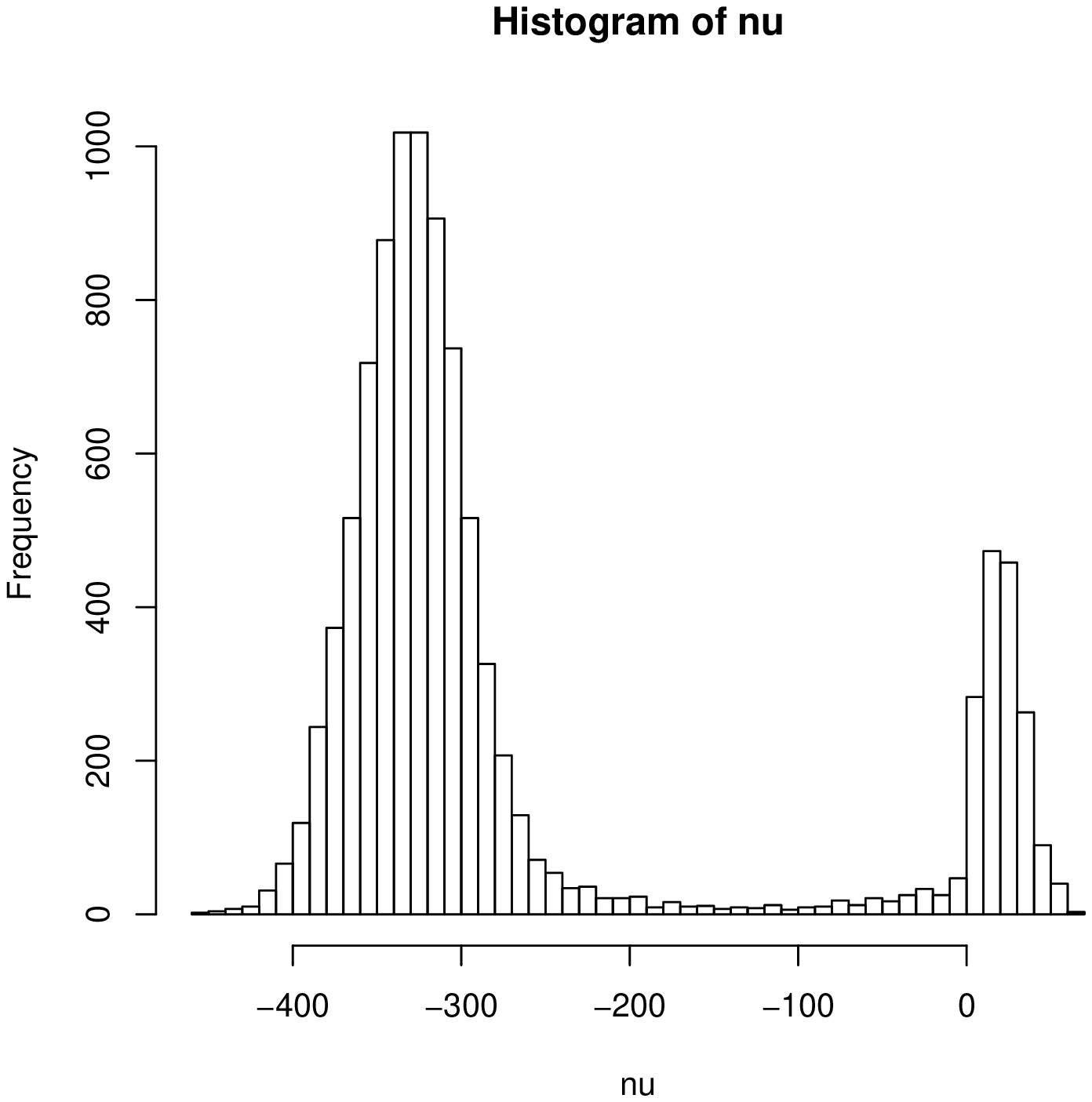,width=0.3\linewidth,clip=} &
\epsfig{file=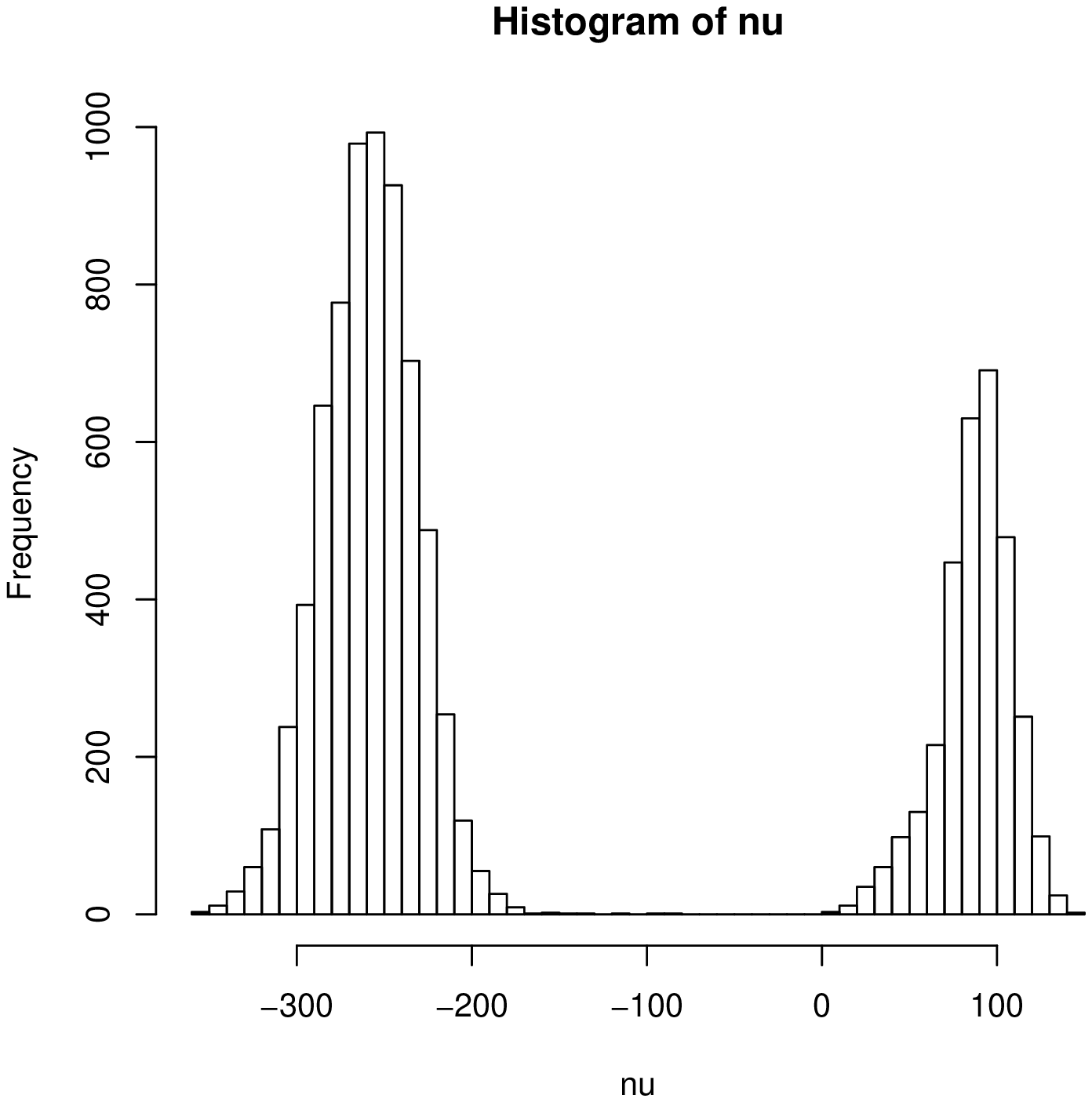,width=0.3\linewidth,clip=}
\end{tabular}
\caption{Distribution of opinions for different amounts of inflexibles in favor of $A$. All simulations started with just a minority supporting $A$, $p_A = 30\%$ and correspond to groups of size 3, where the agents are influenced by their own opinion and only one round of debates happen ($M=1$). In all cases, we had $N=10,000$ agents and $500,000$ random groups were formed. The upper left distribution corresponds to $I_A=0.4$; the upper right, to $I_A=0.6$; lower left, $I_A=0.8$; and lower right to $I_A=0.95$.}\label{fig:self3_propa03}
\end{figure}

Figure (\ref{fig:self3_propa03}) shows the distribution of the strength of opinions, measured by $\nu$s, for groups of size 3, with no initial inflexibles supporting $B$ and initial minority support in favor of $A$, $p_A=30\%$. In those runs, interactions were stopped after just $M=1$ attempt at convincing others and the agent own opinion influences itself. What we see is that if there are not many initial inflexibles supporting $A$, agreement is reached after a while. Even for $I_A=40\%$, the inflexibles are eventually convinced. They are still very distinguishable, forming a second peak with weaker opinions (smaller $|\nu|$ but all agents prefer $B$ ($\nu <0$). As $I_A$ grows, however, the peak is able to survive. For $I_A=60\%$, we still see inflexibles being slowly convinced by the majority, but this effect almost disappears when $I_A=80\%$. For $I_A=95\%$, it is clear that the inflexible peak has moved to the right, with most of its members having an opinion close to $\nu=100$, after starting around 50. That is, both groups are able to reinforce their own point of views, meaning they should survive even in extremely long runs, where their opinions will just keep getting stronger.

\begin{figure}
\centering
\begin{tabular}{ccc}
\epsfig{file=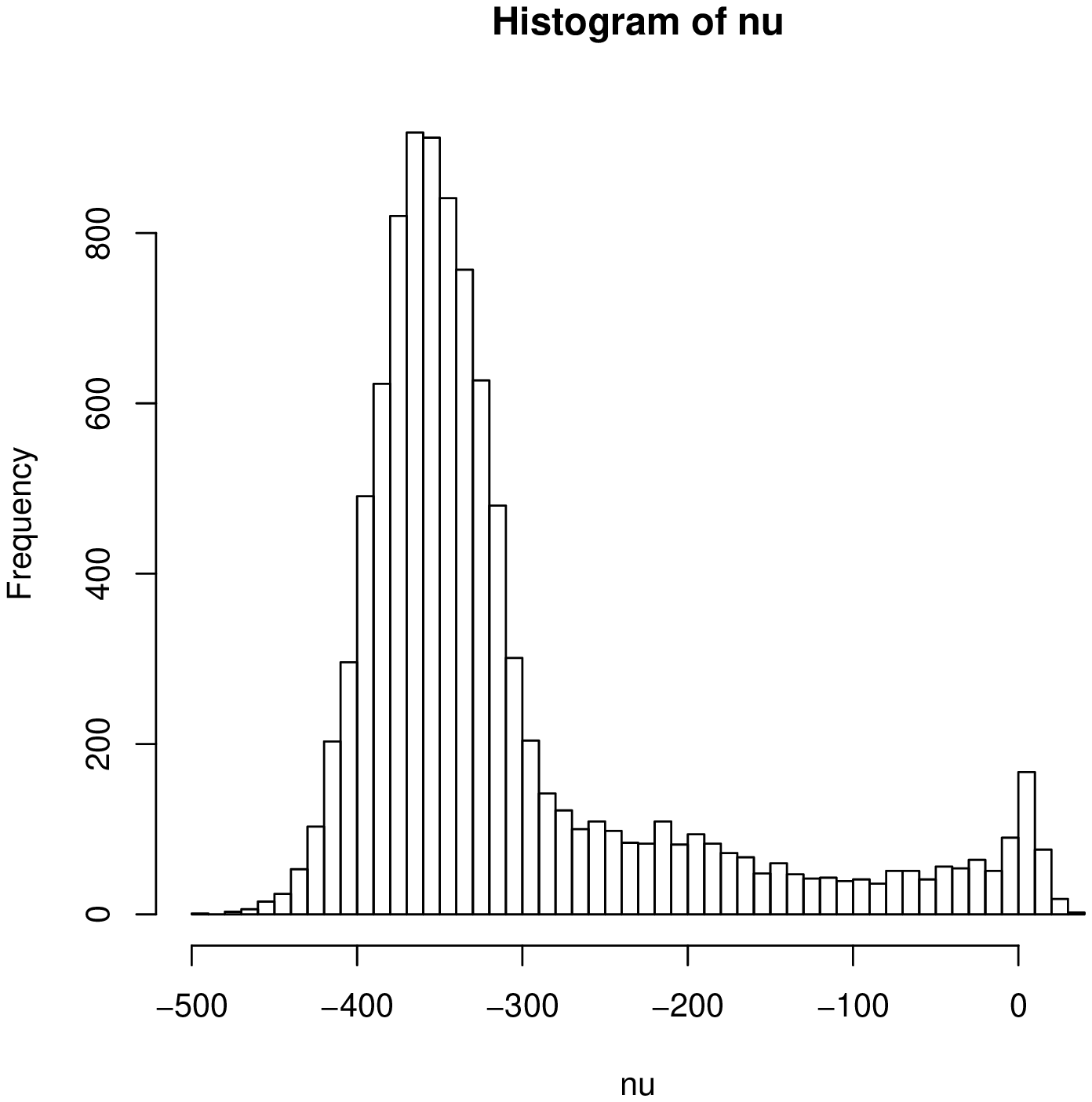,width=0.3\linewidth,clip=} & 
\epsfig{file=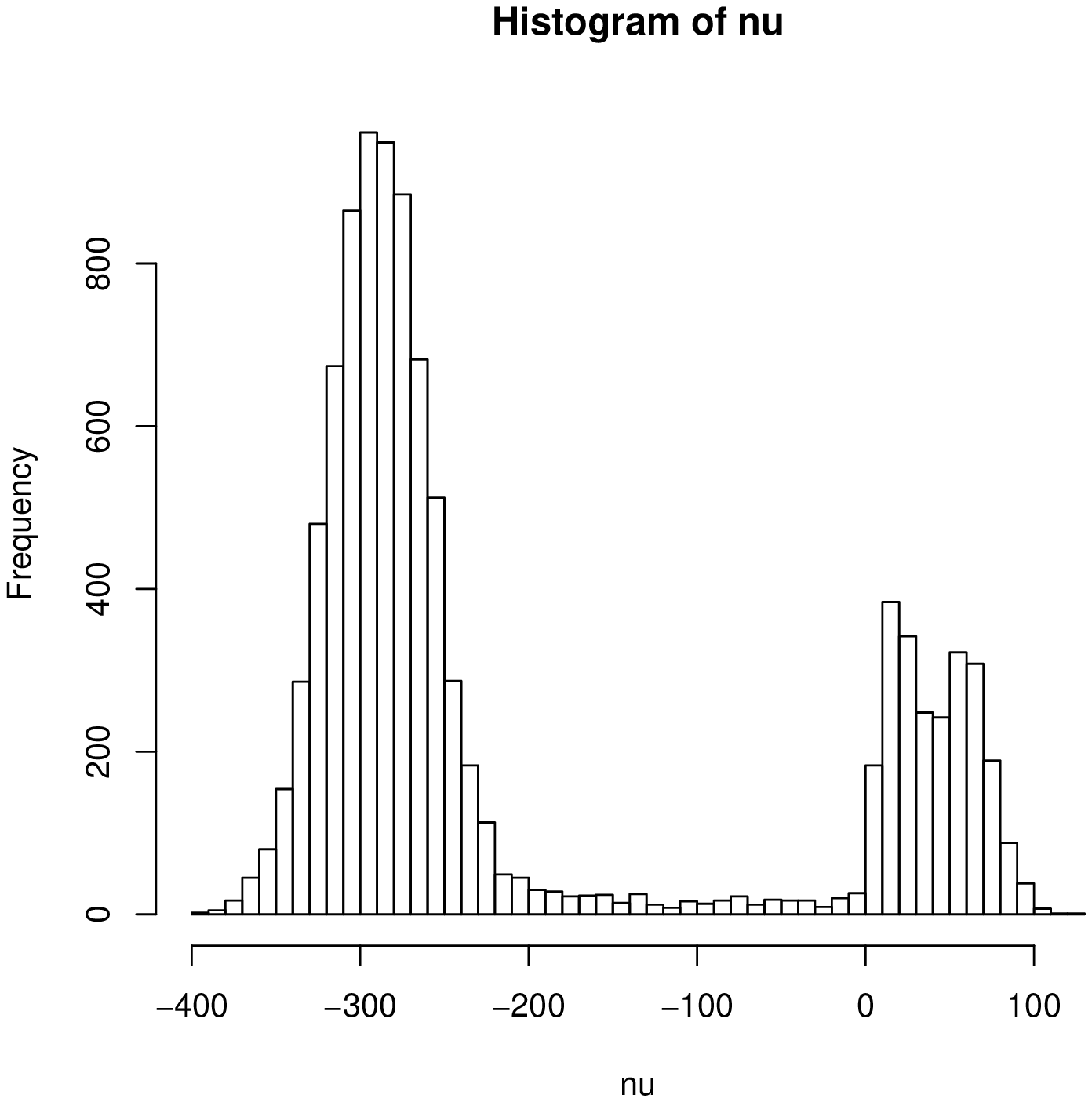,width=0.3\linewidth,clip=} &
\epsfig{file=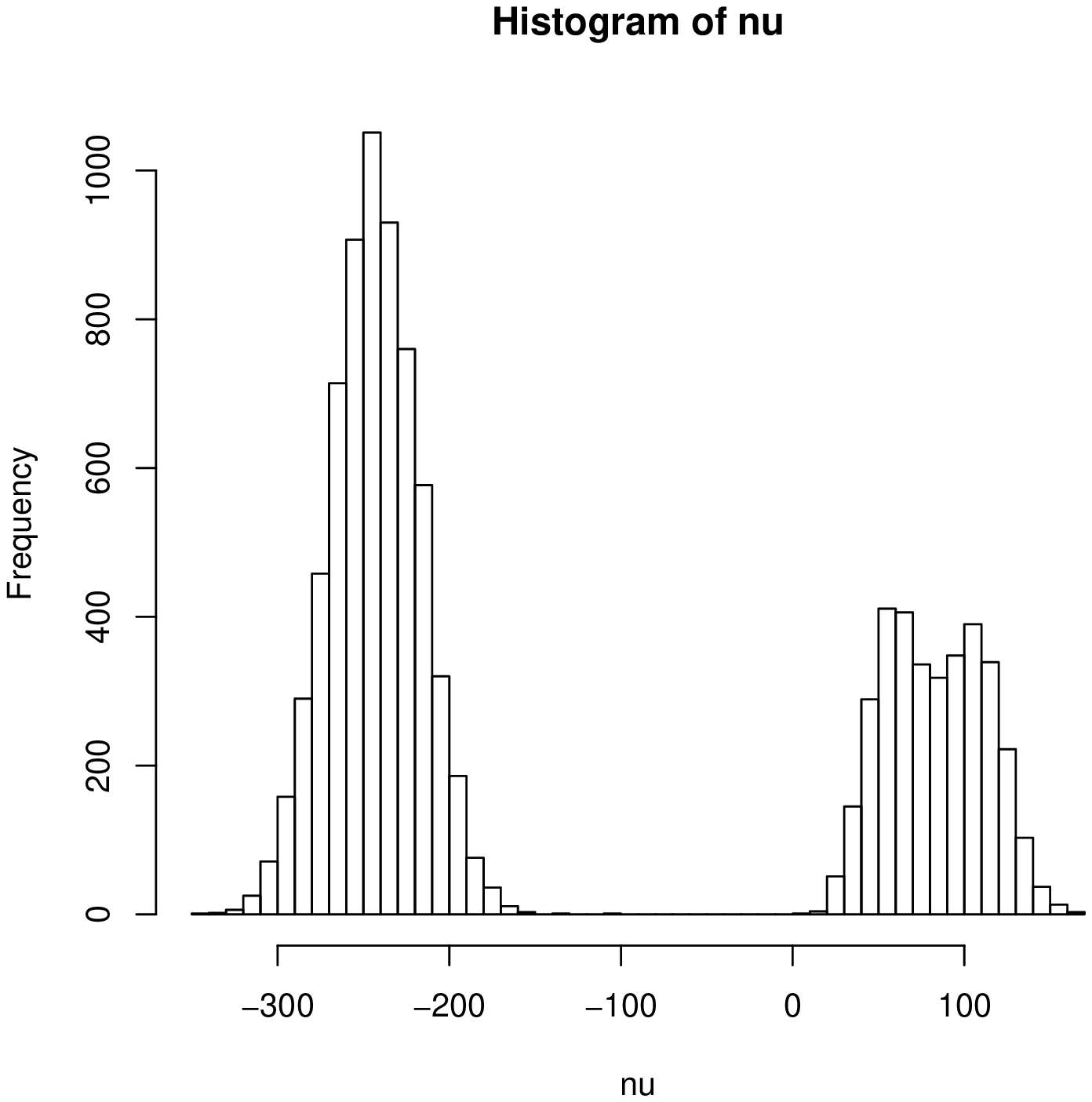,width=0.3\linewidth,clip=}
\end{tabular}
\caption{Distribution of opinions for different amounts of inflexibles in favor of $A$. The figures correspond to the same configuration as in Figure \ref{fig:self3_propa03}, with $N=10,000$, $M=1$, and $500,000$ groups of size 3 formed.  The difference is that the initial minority supporting $A$ is now $p_A = 40\%$. The left distribution corresponds to $I_A=0.2$; central one, to $I_A=0.3$;  and the right one to $I_A=0.4$.}\label{fig:self3_propa04}
\end{figure}

Figure (\ref{fig:self3_propa04}) shows similar results when the proportion of supporters for $A$ is still a minority, but a little larger, $p_A=40\%$. In this case, as expected, a smaller proportion of inflexibles is enough to keep the inflexible peak alive. If  $I_A=20\%$, agreement still happens; at  $I_A=30\%$, we still see a few inflexibles being convinced and it is not clear how much of the inflexible peak would survive in a very long run. But as  $I_A=40\%$, the inflexible peak is clearly reinforcing itself.

We see that under some circumstances, inflexibles can survive by their own influence. In the cases abover, however, they were not observed to be capable of convincing the whole population about their cause, contrary to the results observed in the GUF treatment. What we see here is that, since normal agents get their positions reinforced with each interaction, their opinions eventually become as strong as those of the initial inflexibles and that prevents their group from being convinced by the inflexible initially strong position. Notice that, since self-reinforcement does happen, there is a very strong tendency for everyone to reinforce their views. With just 3 people per group and their own voting counting, only when the other two disagree with the agent will it have its opinion not reinforced. When the other two disagree, the own agent opinion sets the tie in favor of what it already thought, meaning that it really should be expected that agents would tend to keep their initial positions under these circumstances. And, since $M=1$, the opinion can be changed by at most $1$ and that is not enough for the inflexibles to turn the decision.

\begin{figure}
\centering
\begin{tabular}{ccc}
\epsfig{file=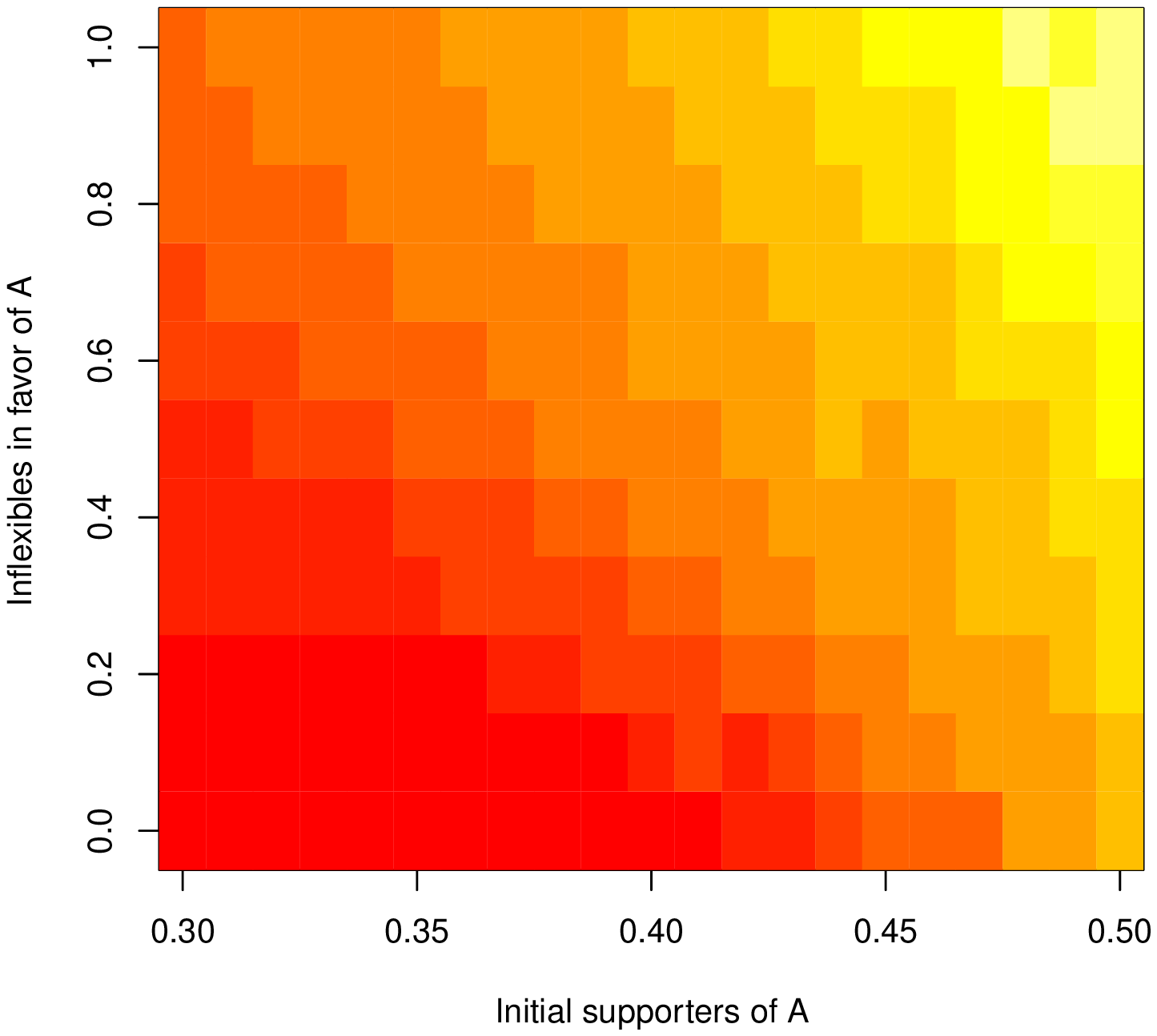,width=0.33\linewidth,clip=} & 
\epsfig{file=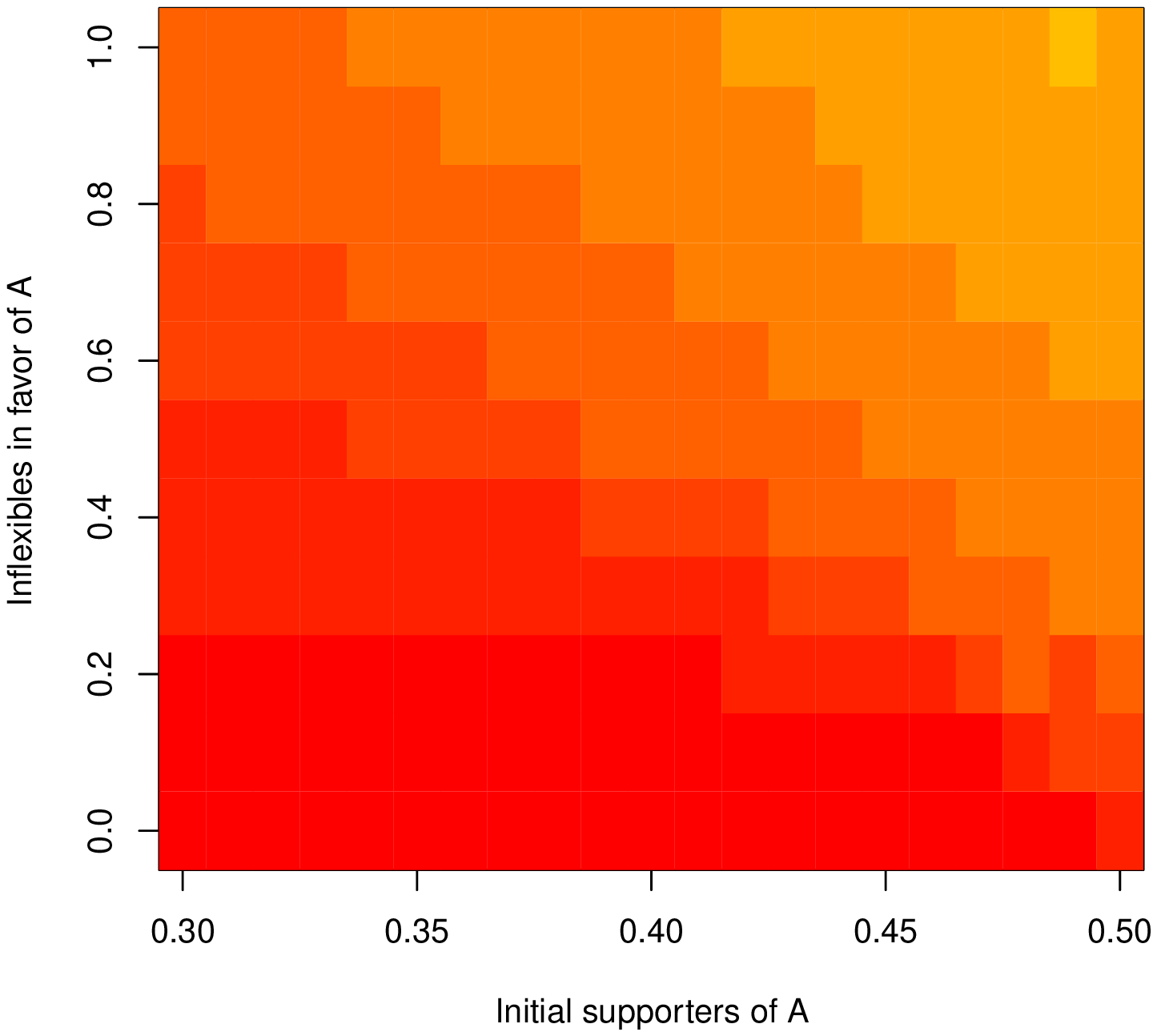,width=0.33\linewidth,clip=} \\
\epsfig{file=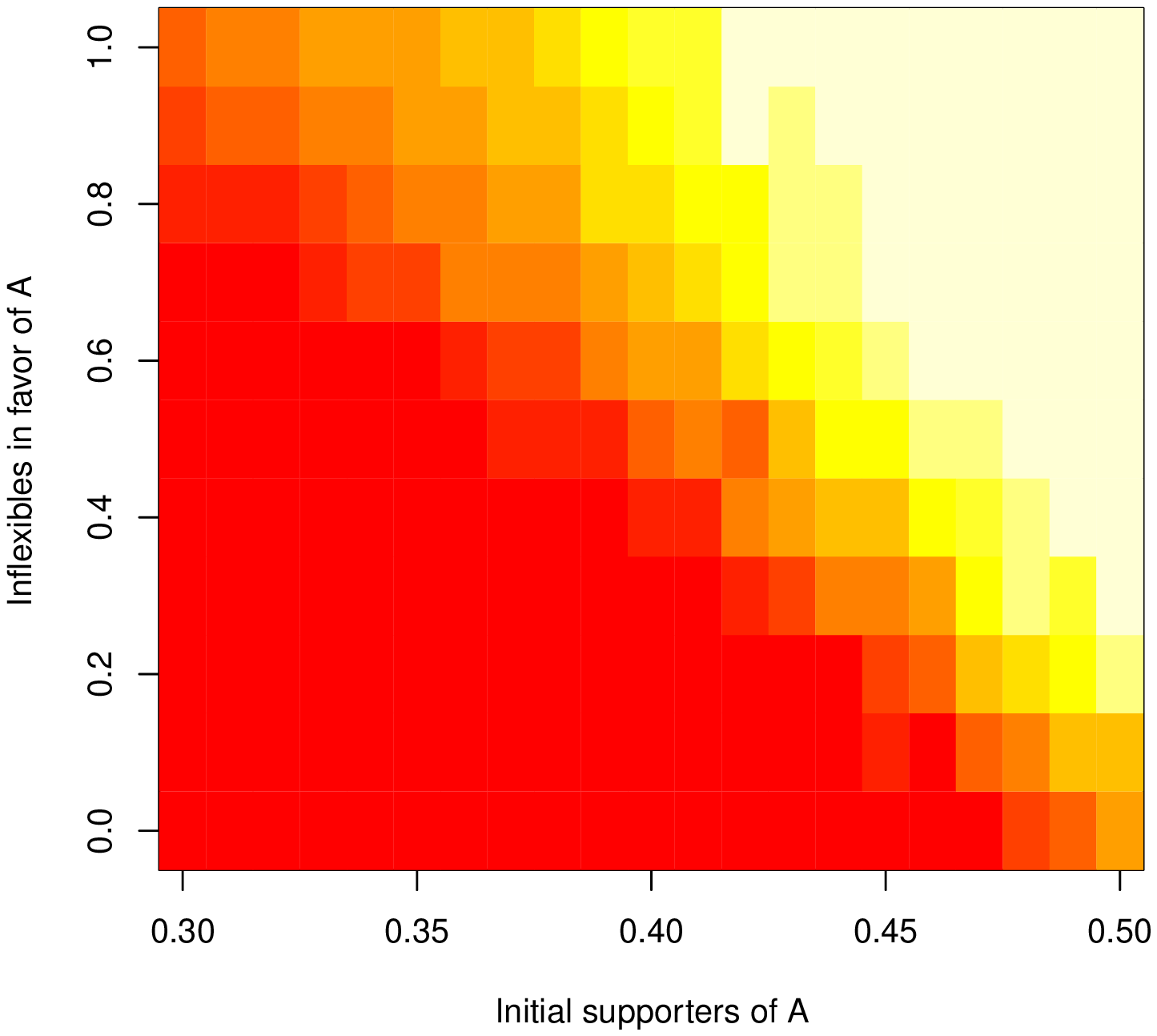,width=0.33\linewidth,clip=} &
\epsfig{file=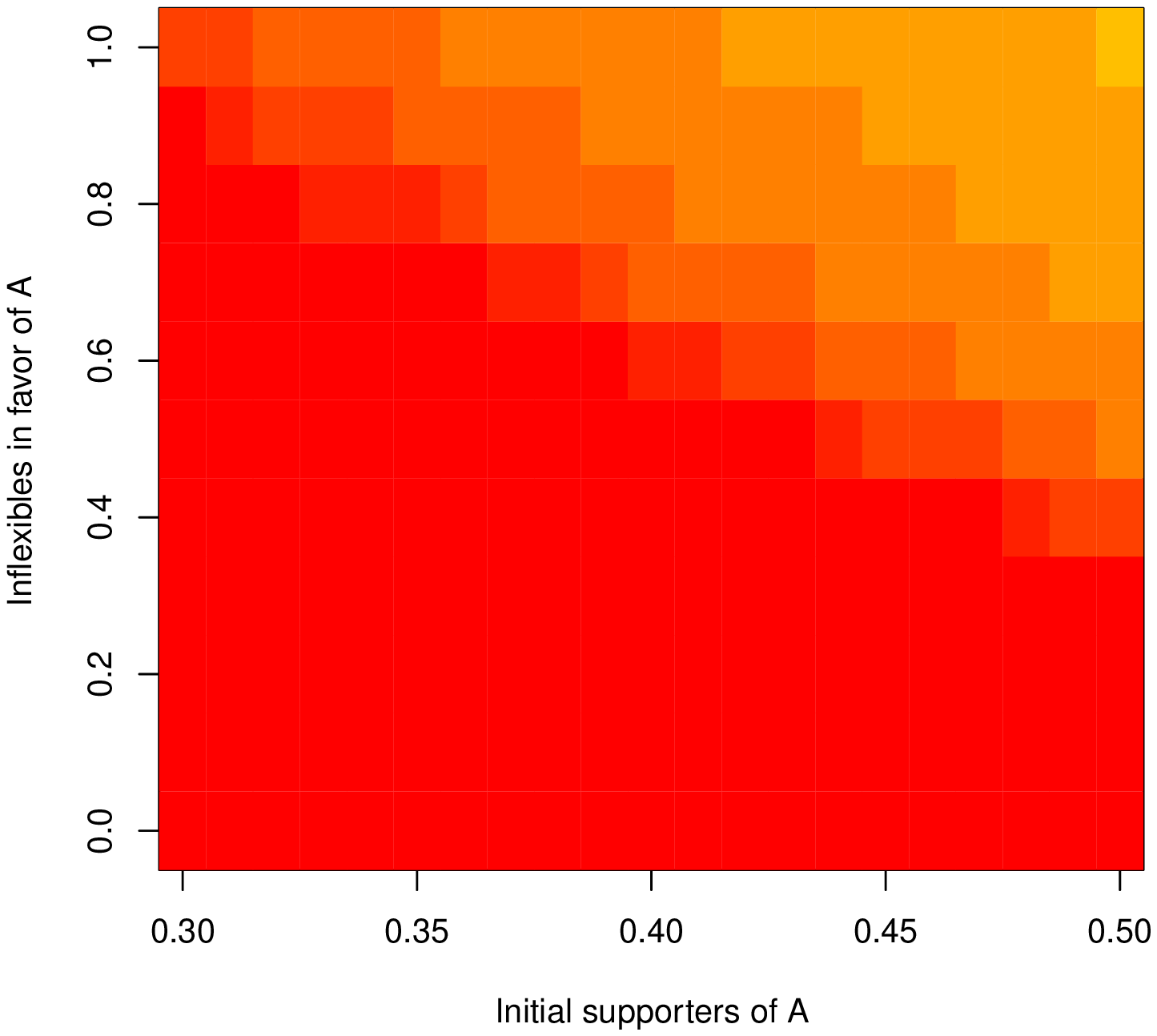,width=0.33\linewidth,clip=}\\
\epsfig{file=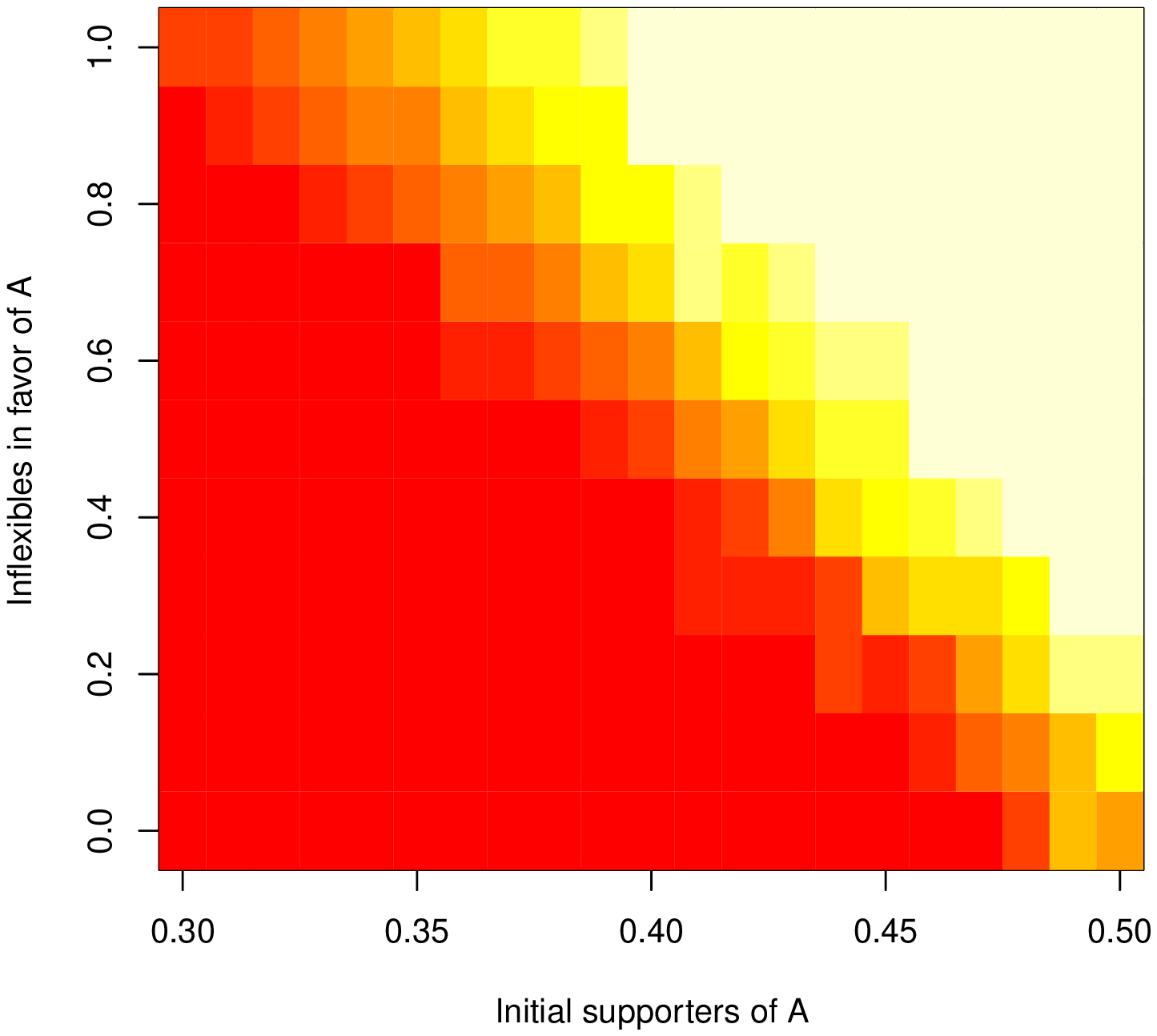,width=0.33\linewidth,clip=} &
\epsfig{file=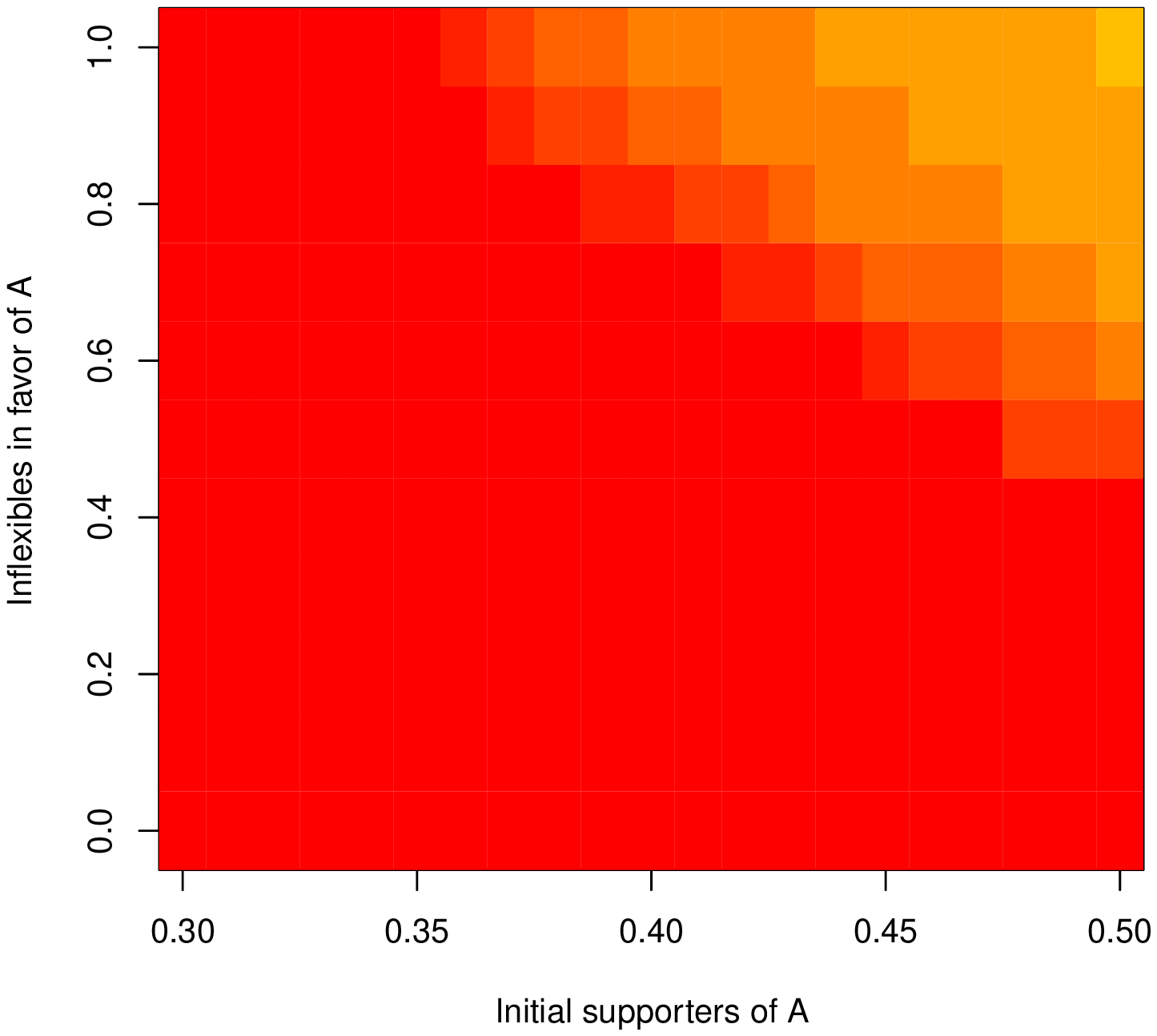,width=0.33\linewidth,clip=}
\end{tabular}
\caption{Average over 20 realizations of the final proportion of agents supporting $A$ for different values of the parameters, as a function of both $p_A$ and $I_A$, where the agent is influenced by its own opinion. Red correspond to a final state where all agents chose $B$ and white, to $A$ being chosen by all; tones of yellow show the intermediary cases. The top line figures correspond to $M=1$, the middle one, to $M=5$, and the bottom one to $M=9$. Left most figures show results where $I_B=0$,  and the right ones, to $I_B=10\%$ }\label{fig:self3aver}
\end{figure}

In the discrete case of GUF, given the right initial parameters, the majority always convinced the minority. In order to get closer to that result, it makes sense to make $M$ larger so that, if the minority agent is not convinced at first, debates go on for a while, allowing for stronger influence effects. Results for different values of $M$ can be seen in Figure  (\ref{fig:self3aver}). There we see that, for $M=1$, even for larger proportions of $I_A$, the population never shifts to a belief in $A$, that would be represented by the white color. For small $I_A$ and $p_A$ (down left at each graphic), obviously, the whole population gets convinced that $B$ is the best option (represented by the red color in the graphics). As both inflexibles and initial supporters of $A$ become more common, we can see that, even for $M=1$, it is no longer possible for the majority supporting $B$ to convince everyone. But the minority also fails to convince the majority.

The behavior is clearly different when $M=5$ or 9. For larger values of $M$, an white area appears, meaning that the initial majority of $B$ supporters is convinced by the inflexibles and the system is changed into an agreement where everyone actually chooses $A$.
It is important to notice, since the Figure shows values that are averaged over 20 realizations, that the large region where both supporters of $A$ and $B$ (especially $M=1$ and $I_B=0$) are observed corresponds to actual coexistence of both choices in each run and not different final results where each realization would have ended with al agentsl in favor of a single, but each time different, choice.

\subsubsection{The agent self-exclusion case}
It makes sense now to inquire what happens if the agent does not count itself towards the majority.  We have then that a few differences in the behavior of the system can be observed when no self interaction is allowed. In that case, the opinion of the rest of the individuals has a much better chance to influence the agent and we see that, after a while, agreement was observed in all realizations of the model. The important question that remains is to identify how likely it is for a system to end up supporting $A$ or $B$. 
\begin{figure}
\centering
\begin{tabular}{cc}
\epsfig{file=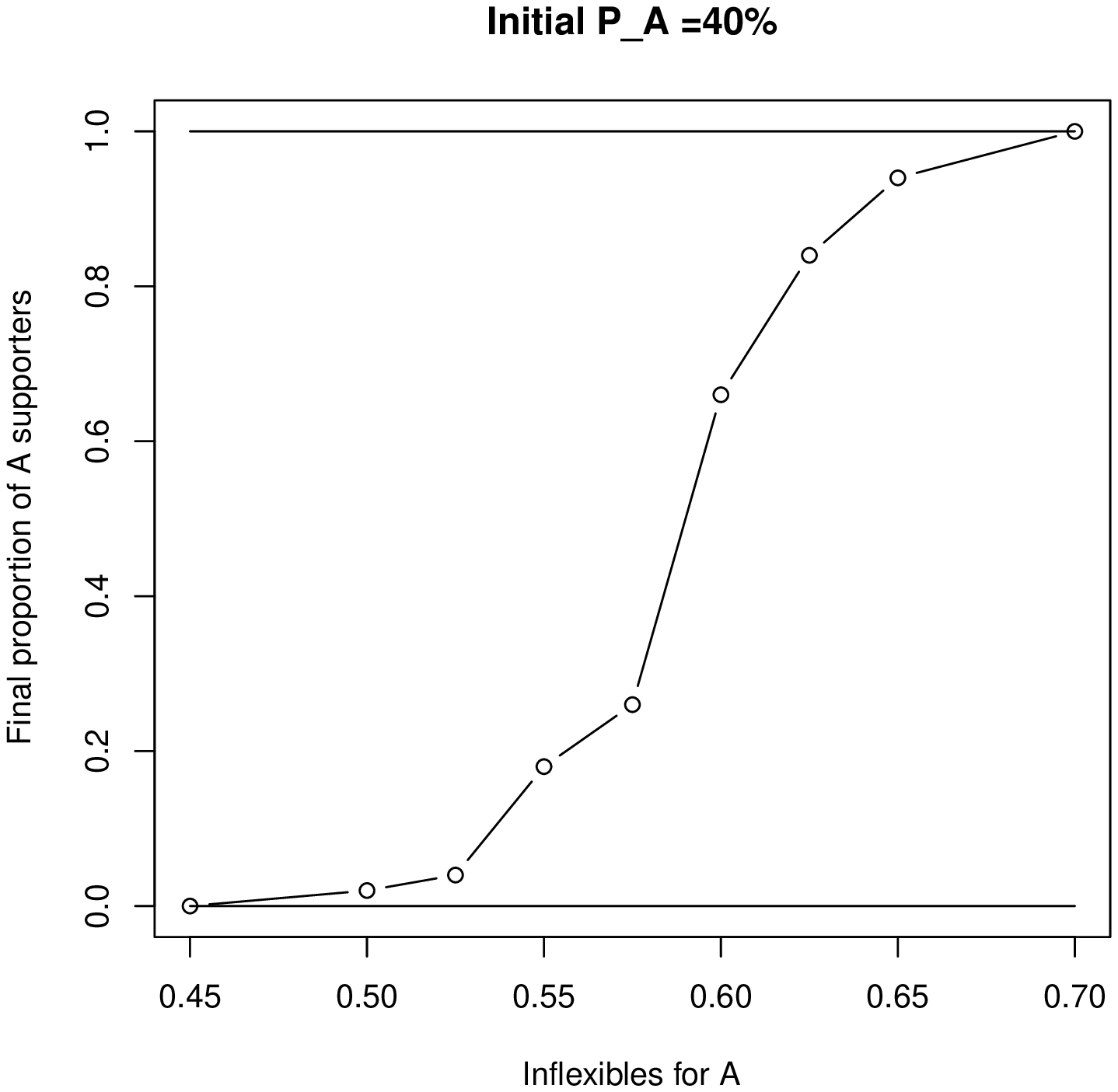,width=0.45\linewidth,clip=} & 
\epsfig{file=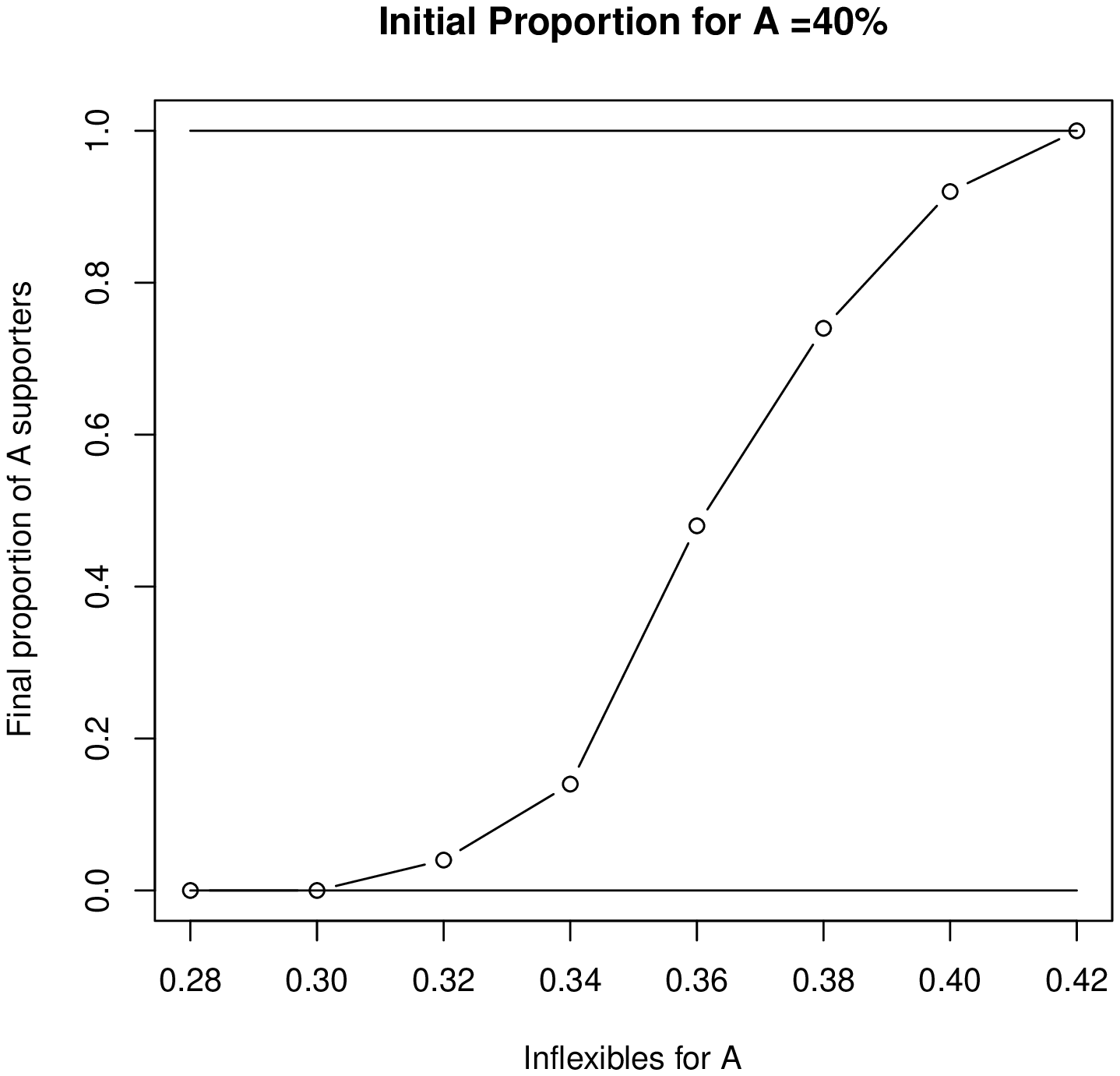,width=0.45\linewidth,clip=}
\end{tabular}
\caption{Proportion of runs after 50 realizations that ended with full agreement in favor of $A$ as a function of $I_A$. In all realziations, initially, there was a 40\% proportion of $A$ supporters. The graphic at the left corresponds to $M=1$ and the one at right to $M=5$.}\label{fig:noself3}
\end{figure}

Simulations to answer how much the initial inflexibles can have in an influence in this scenario are shown in Figure(\ref{fig:noself3}). Each point there corresponds to a proportion obtained over 50 realizations; the left panel shows what happens when agents interact only once in each debate ($M=1$), the right panel shows the results when they try up to five rounds of debates before giving up the attempt to come to an  agreement ($M=5$). We can see that in both cases the inflexibles can turn the tide of a majority in favor of $A$ (initial $p_A$ was 40\% in both cases). Even for $M=1$, it is now possible for a strong minority to convince the majority of their point of view. The observed continuous curve can be an effect of finite size conditions. What we observe is that if the groups are more insistent in reaching some agreement, the influence of the inflexibles grow, as less of them are required to convince the whole population about their claim. It is interesting to remember that the limit $M\rightarrow \infty$ leads us to the discrete case.

\begin{figure}
\centering
\begin{tabular}{ccc}
\epsfig{file=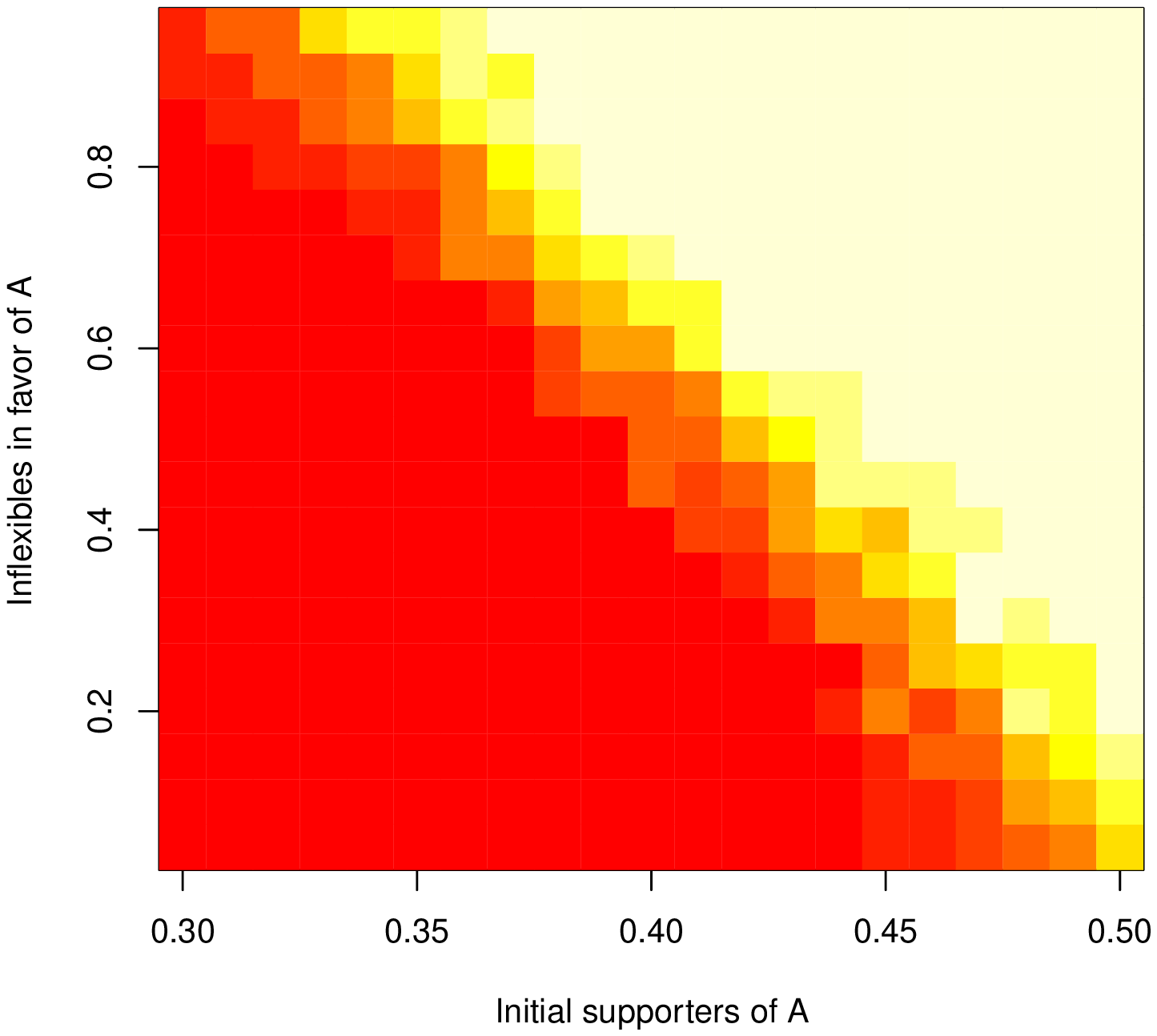,width=0.33\linewidth,clip=} & 
\epsfig{file=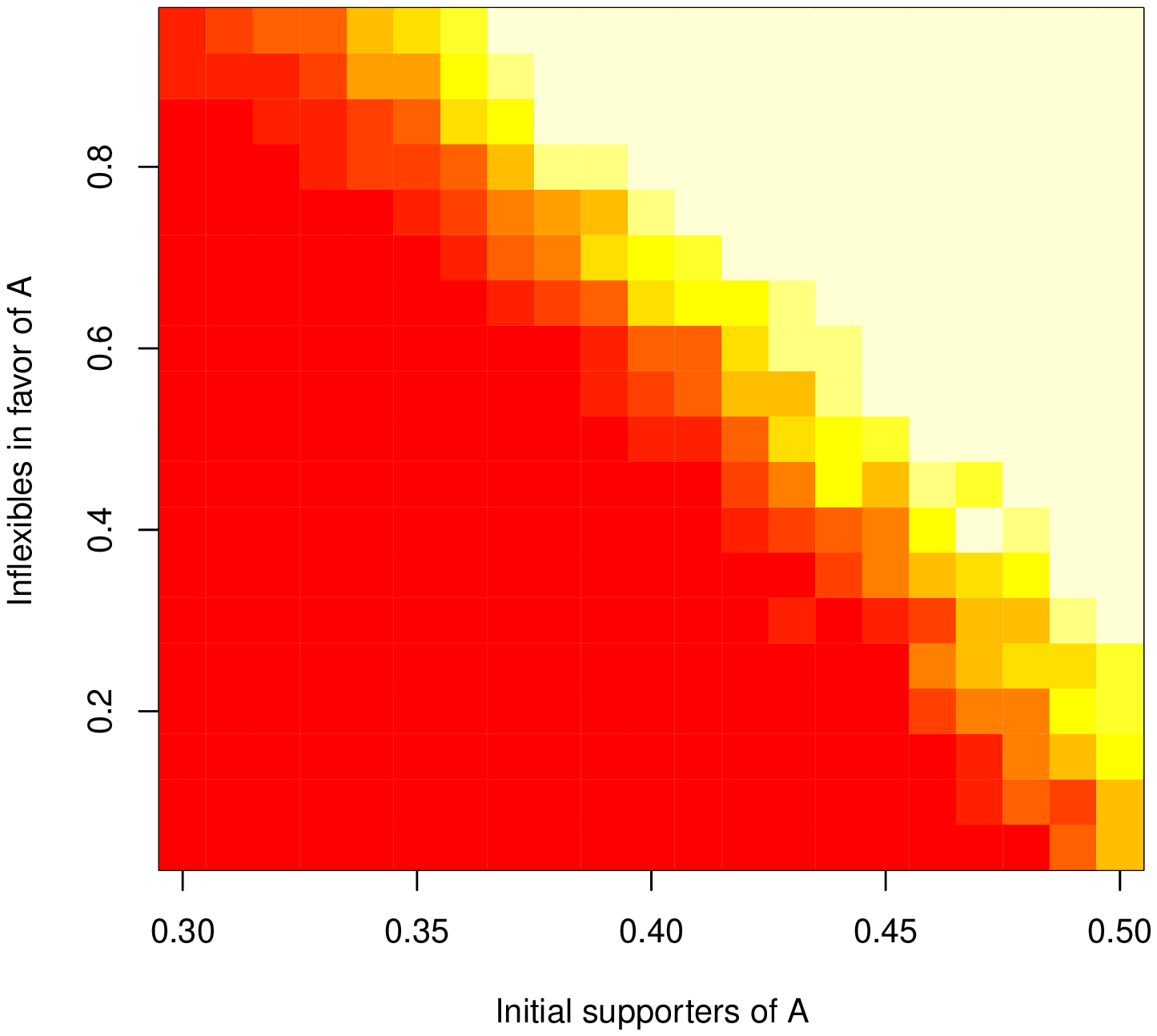,width=0.33\linewidth,clip=} \\
\epsfig{file=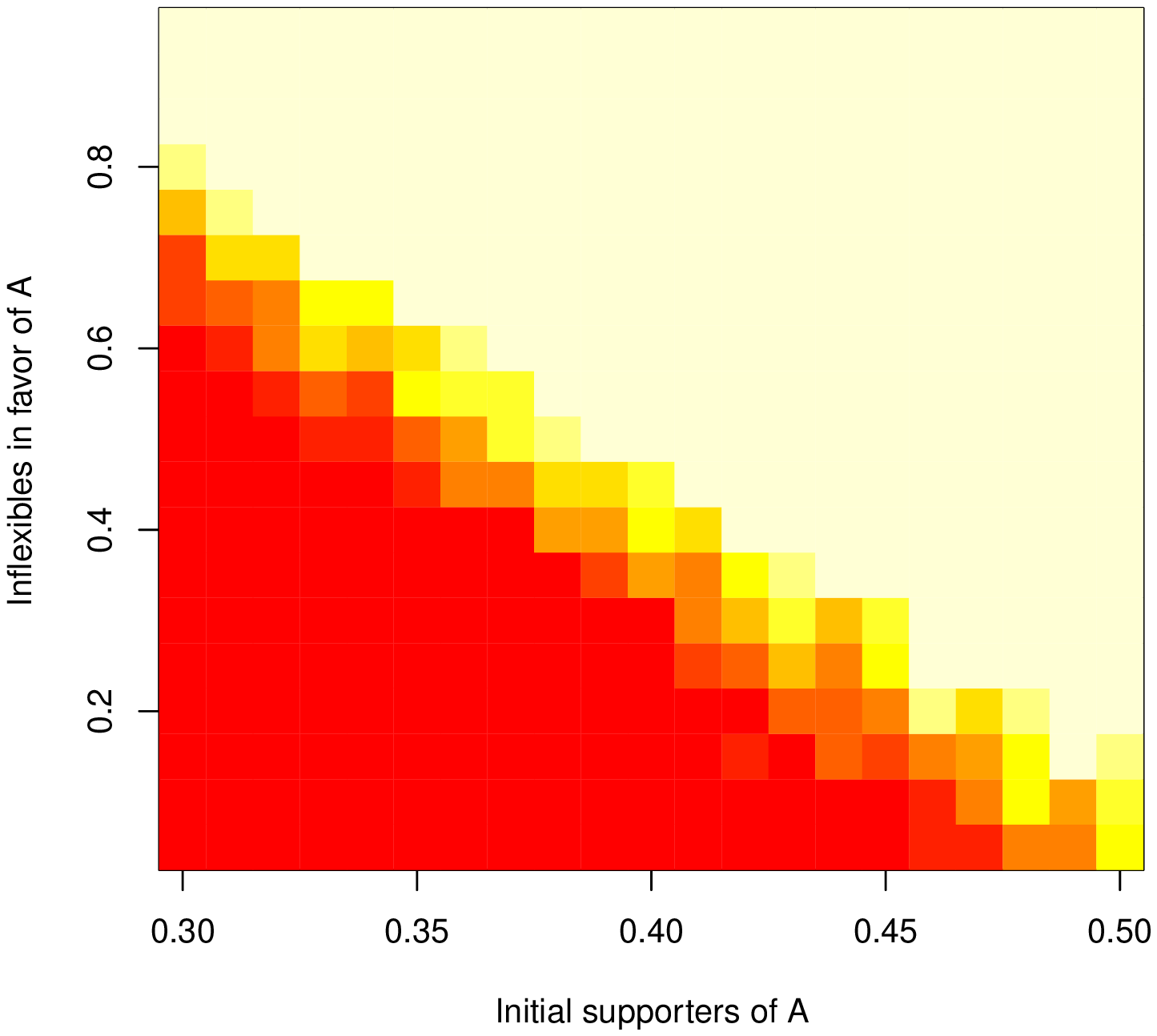,width=0.33\linewidth,clip=} &
\epsfig{file=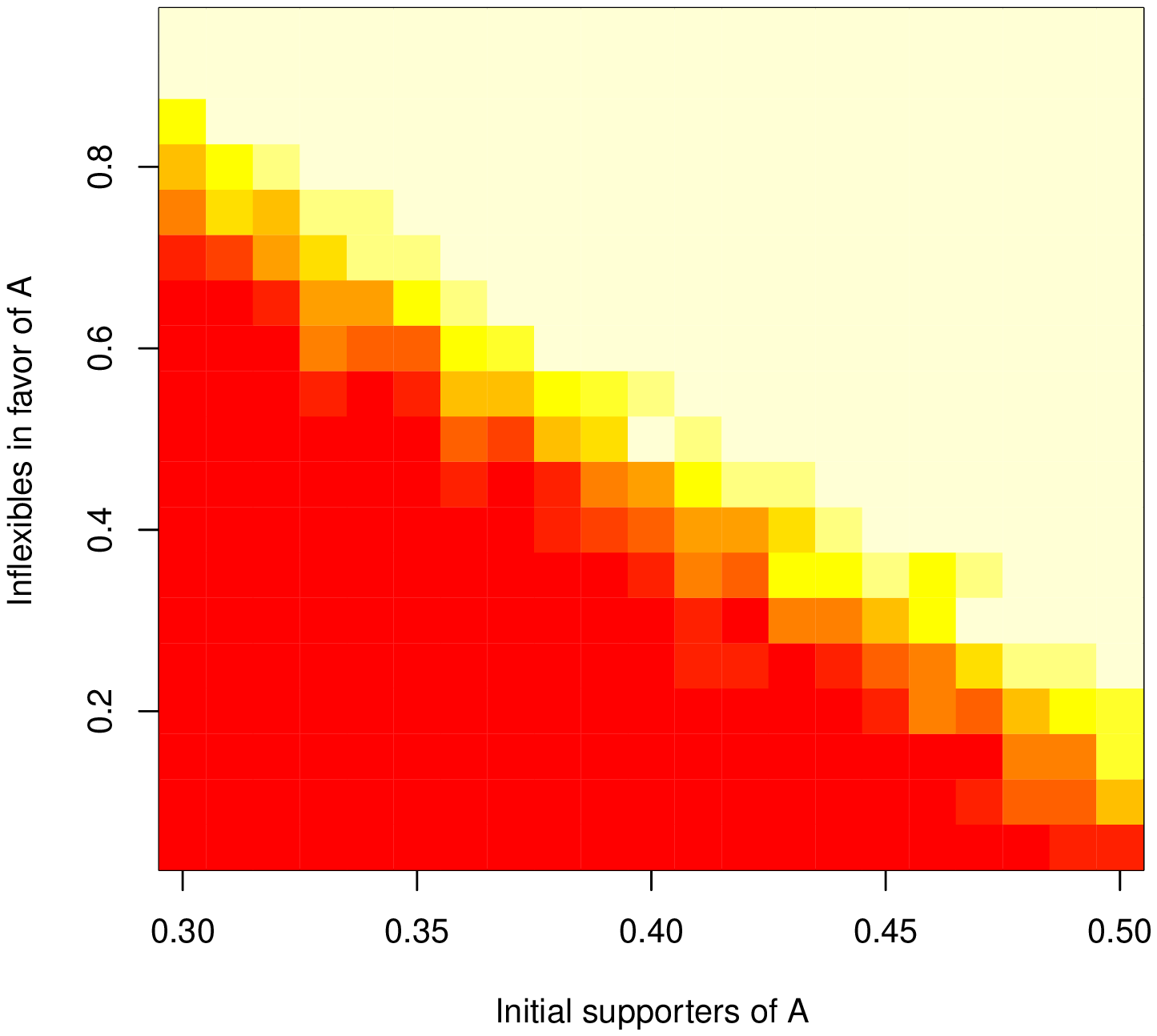,width=0.33\linewidth,clip=}
\end{tabular}
\caption{Average over 20 realizations of the final proportion of agents supporting $A$ for different values of the parameters, as a function of both $p_A$ and $I_A$, where the agent is not influenced by its own opinion. Red correspond to a final state where all agents chose $B$ and white, to $A$ being chosen by all; tones of yellow show the intermediary cases. The top line figures correspond to $M=1$ and the bottom one to $M=5$. Left most figures show results where $I_B=1\%$,  and the right ones, to $I_B=10\%$ }\label{fig:nonself3aver}
\end{figure}

The average results for the same condition (no self interaction in groups of size 3) can also be seen in Figure (\ref{fig:nonself3aver}). It is clear to see the regions where agreement is reached (red for $B$ and white for $A$), as well as the transition regions, where different runs ended with agreements for $A$ or $B$, with the proportions represented by the different tones of orange and yellow between red and white. As a check of consistency, small amounts of inflexibles in favor of $B$ were introduced, with the expected result that they make it harder for the inflexibles in favor of $A$ to convince the whole population.

\subsection{Size 4 groups}

All results presented so far were for groups of size 3. Of course, people can debate in larger groups and it is worth exploring the differences this may introduce. Therefore, in this Subsection, we study groups with 4 people in each. Simulations were run for different values of all the major variables ($p_A$, $I_A$, $I_B$, and $M$), for both the case with self influence and the one without it.

\begin{figure}
\centering
\begin{tabular}{ccc}
\epsfig{file=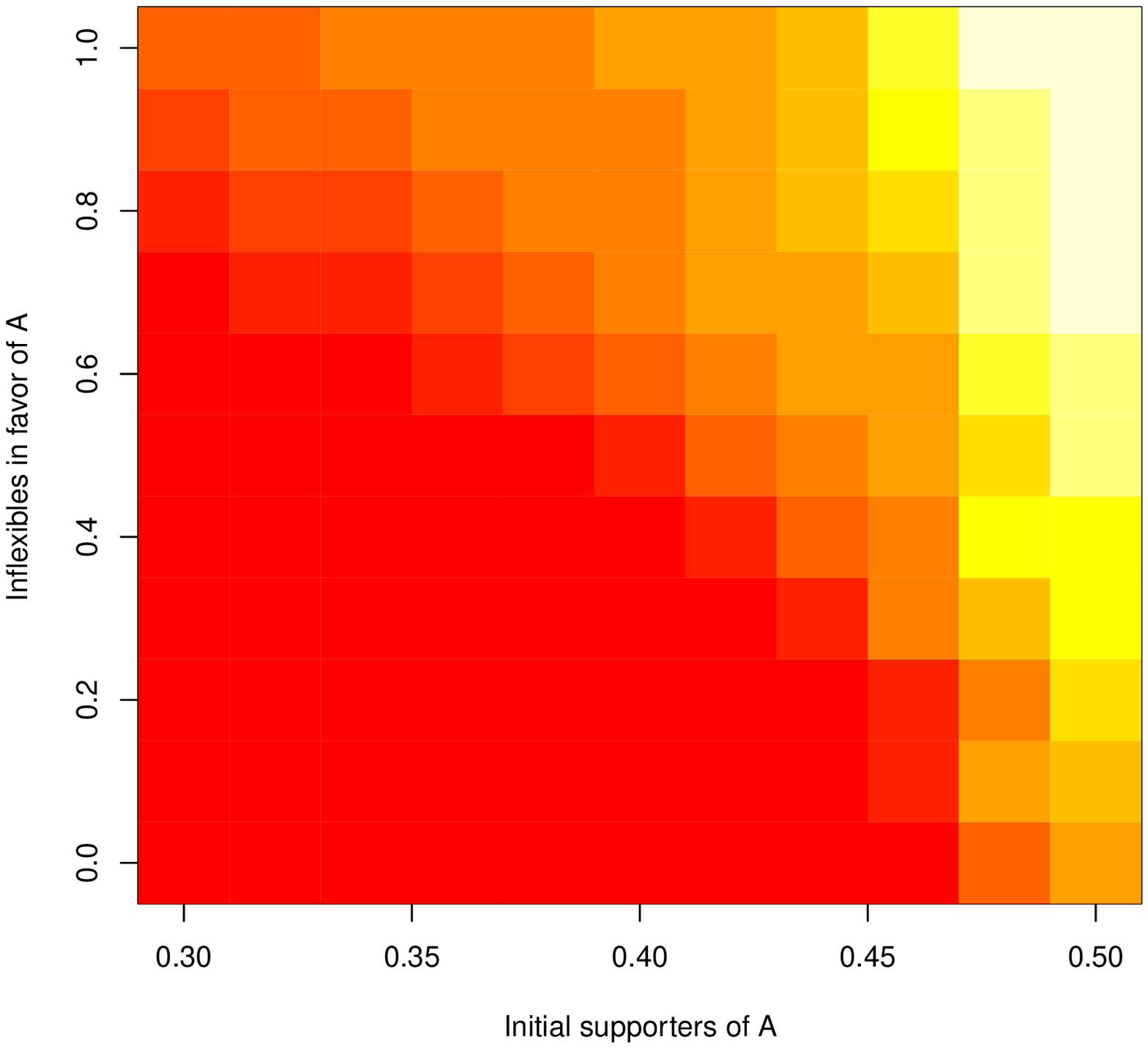,width=0.28\linewidth,clip=} & 
\epsfig{file=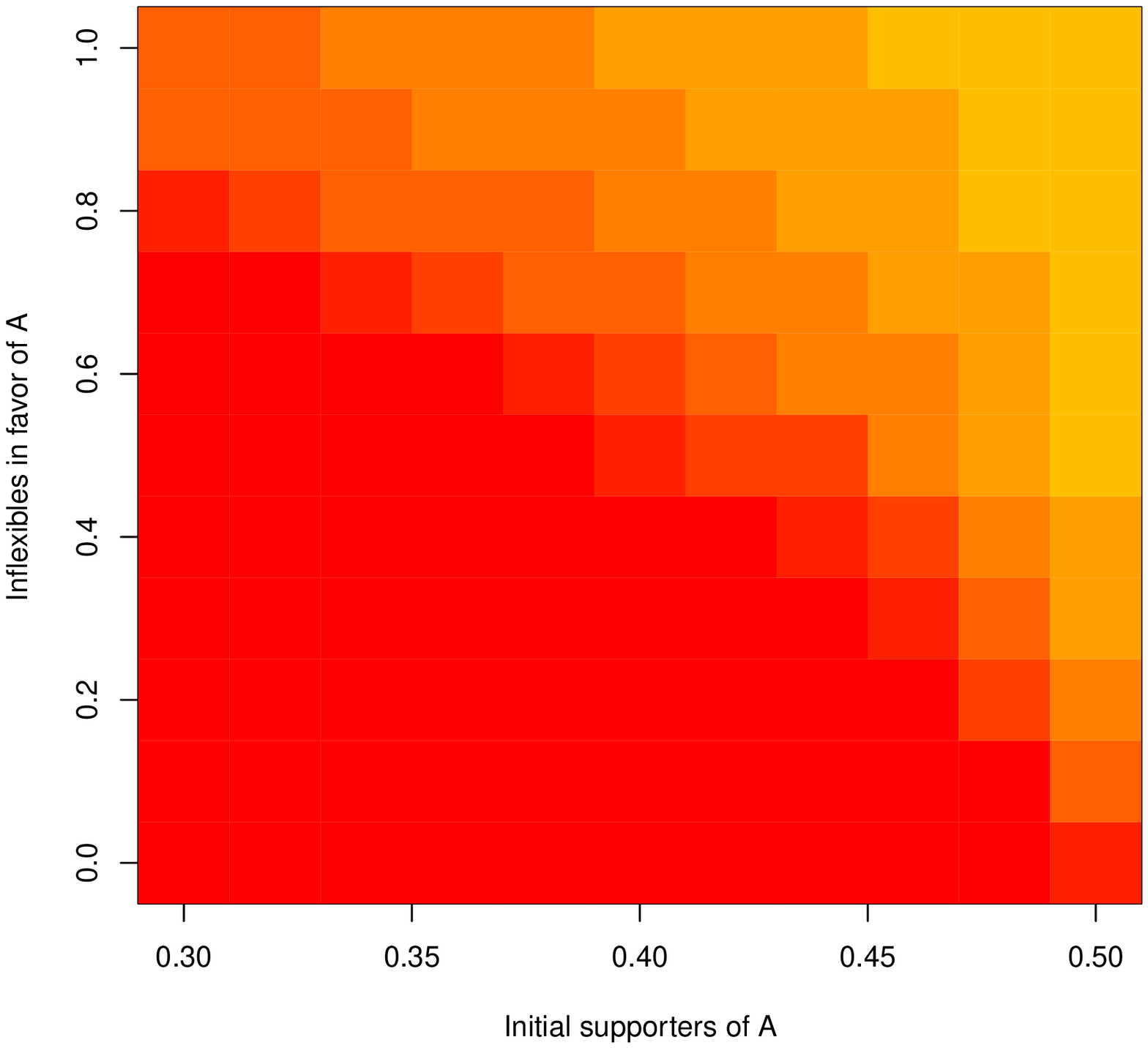,width=0.28\linewidth,clip=} & 
\epsfig{file=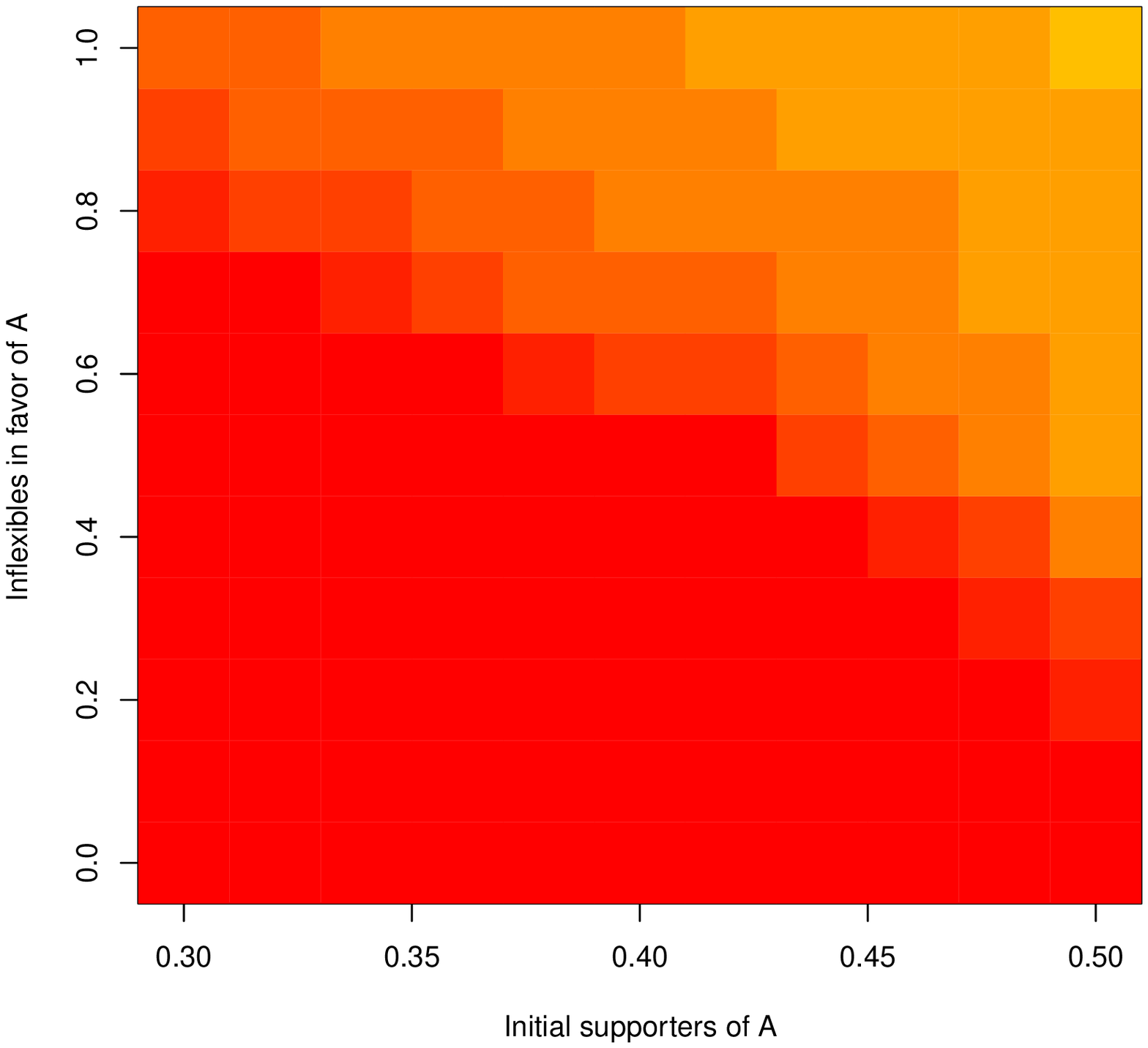,width=0.28\linewidth,clip=} \\
\epsfig{file=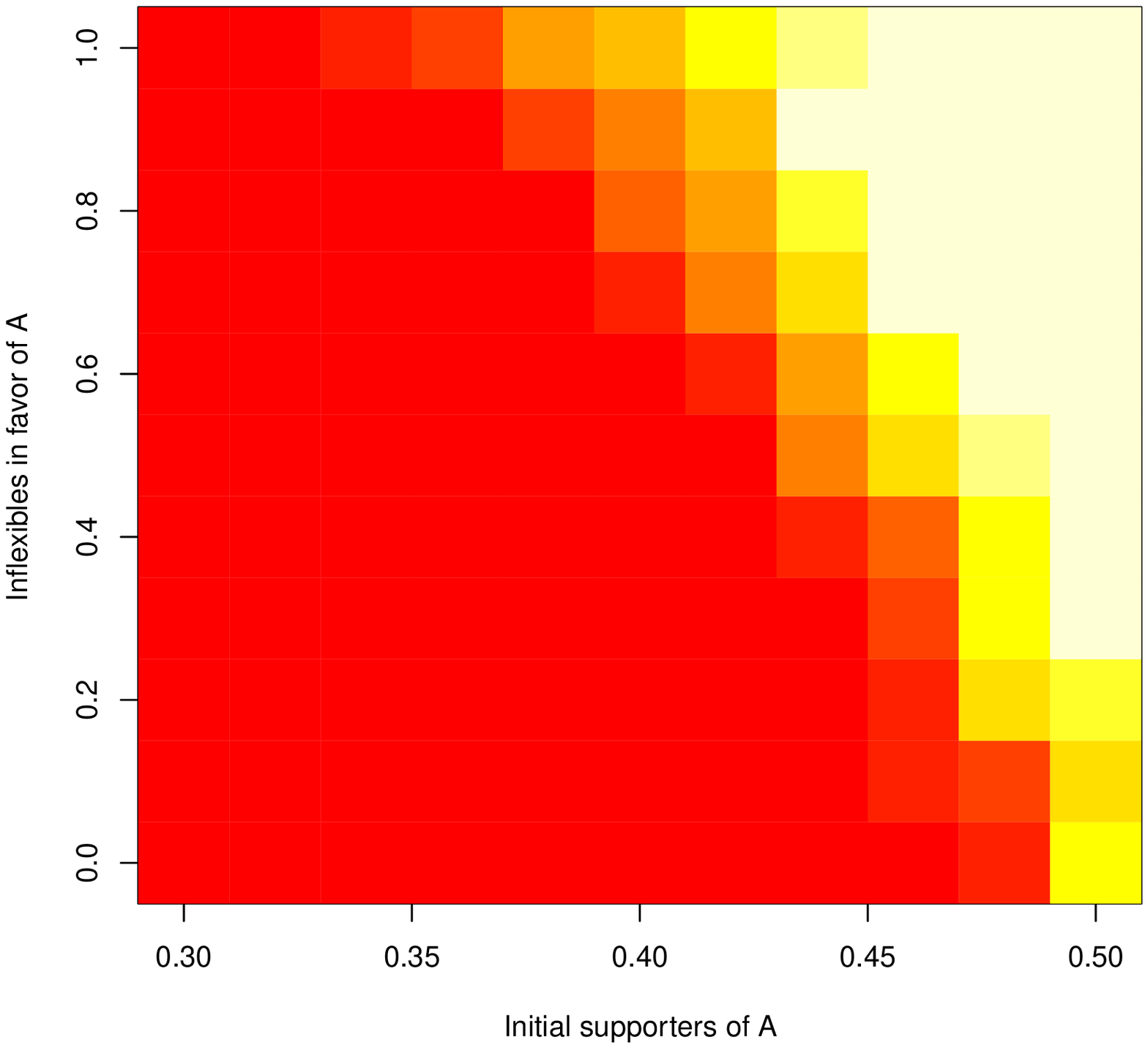,width=0.28\linewidth,clip=} & 
\epsfig{file=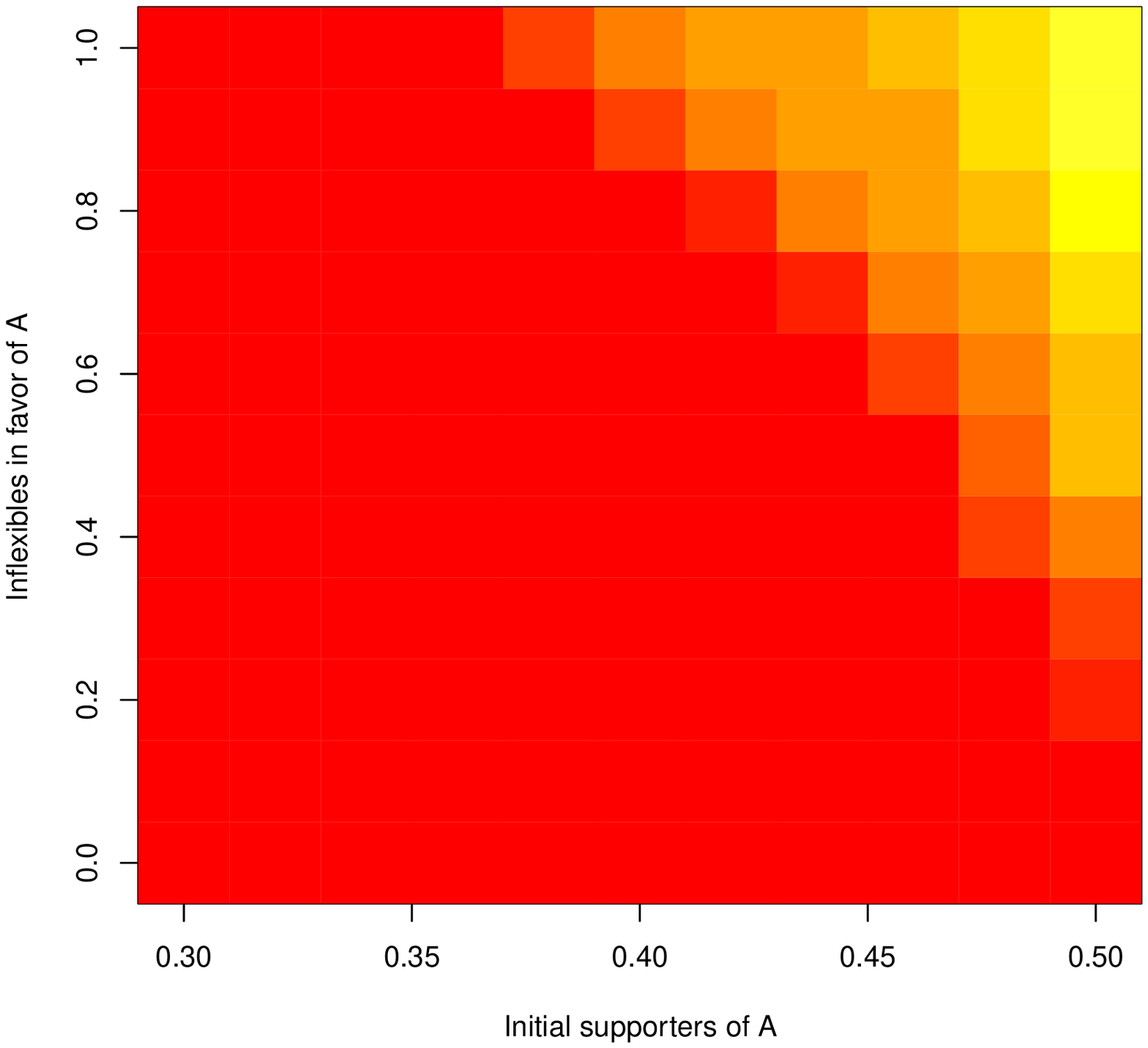,width=0.28\linewidth,clip=} & 
\epsfig{file=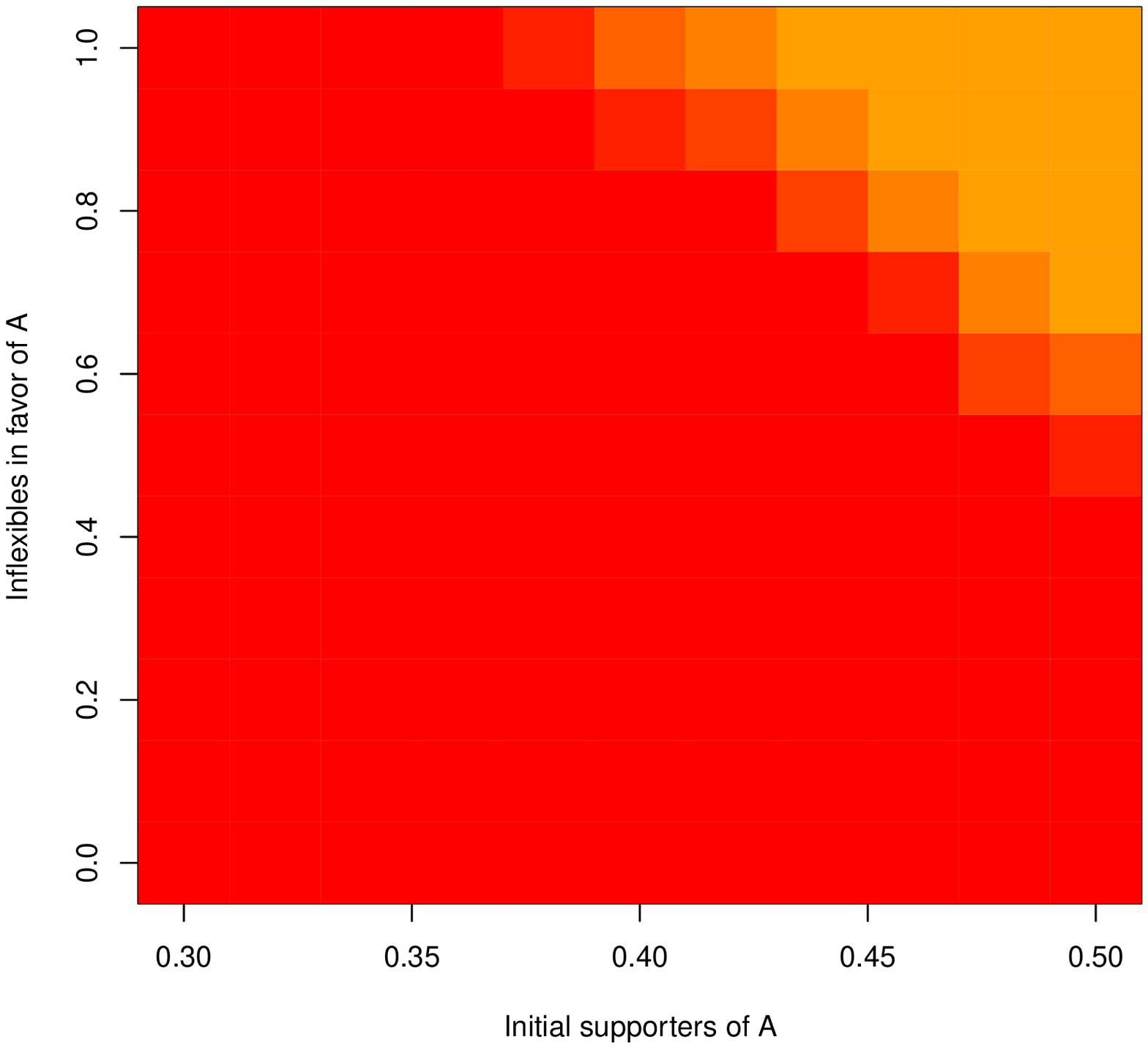,width=0.28\linewidth,clip=}\\
\epsfig{file=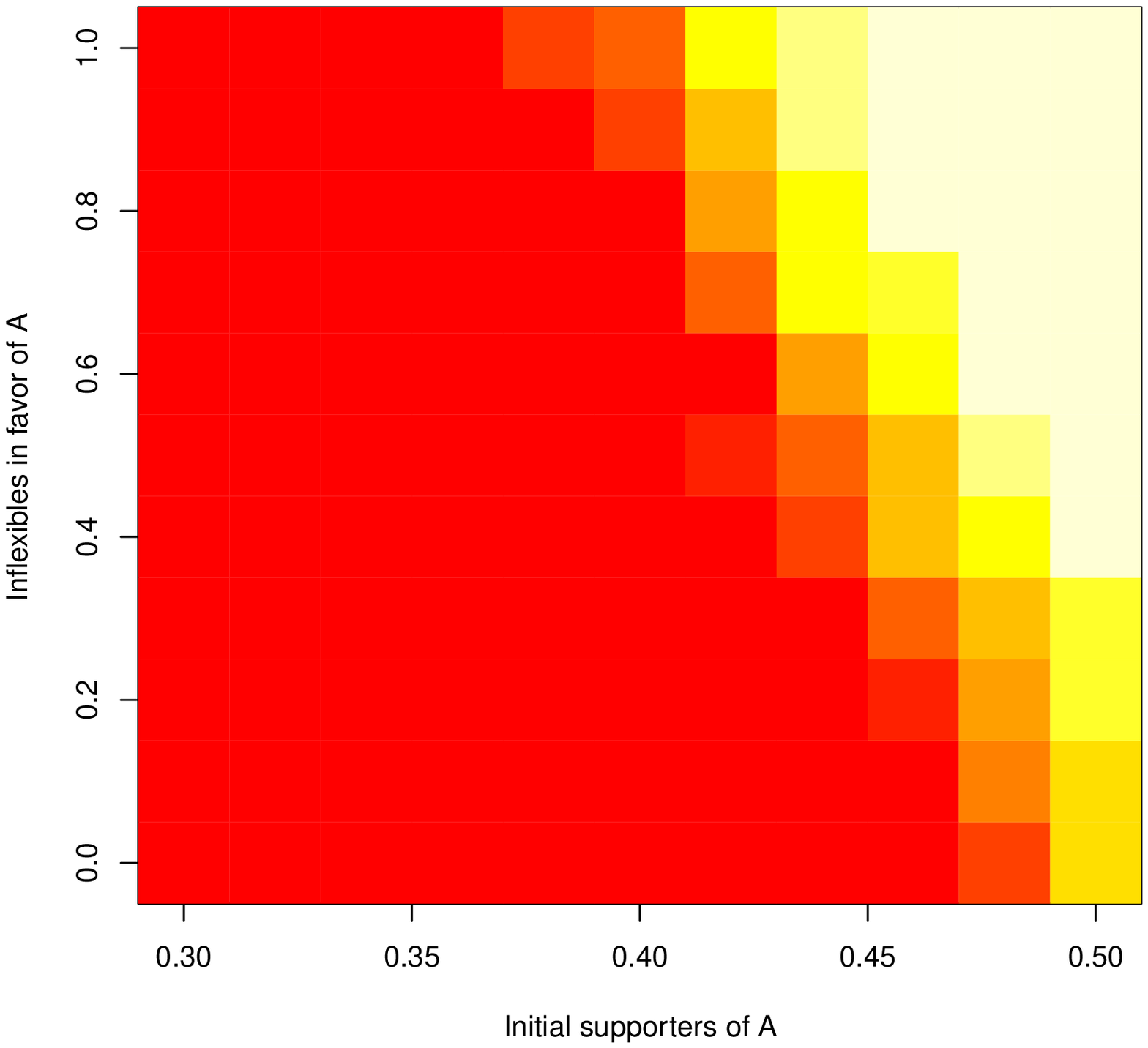,width=0.28\linewidth,clip=} & 
\epsfig{file=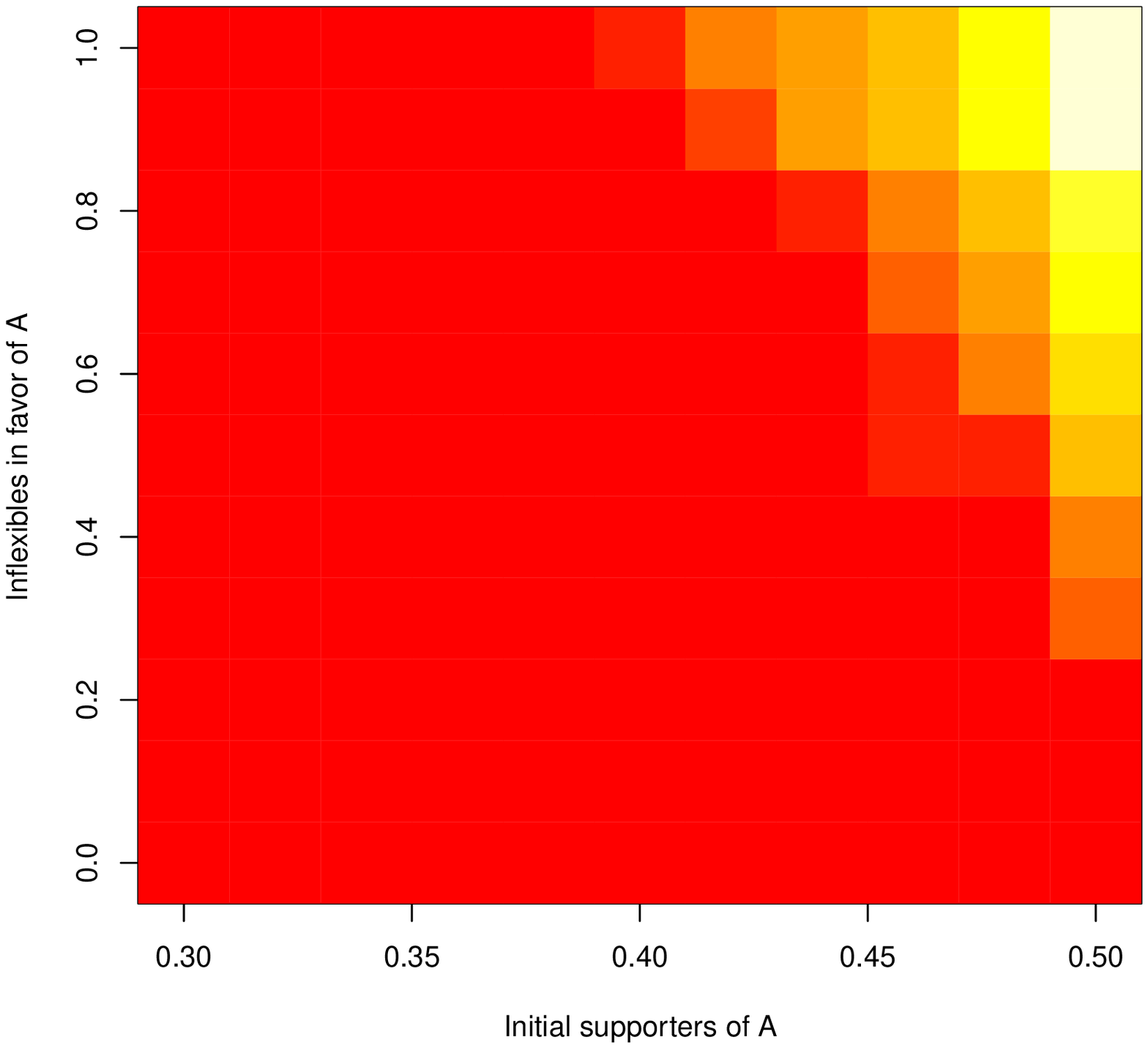,width=0.28\linewidth,clip=} & 
\epsfig{file=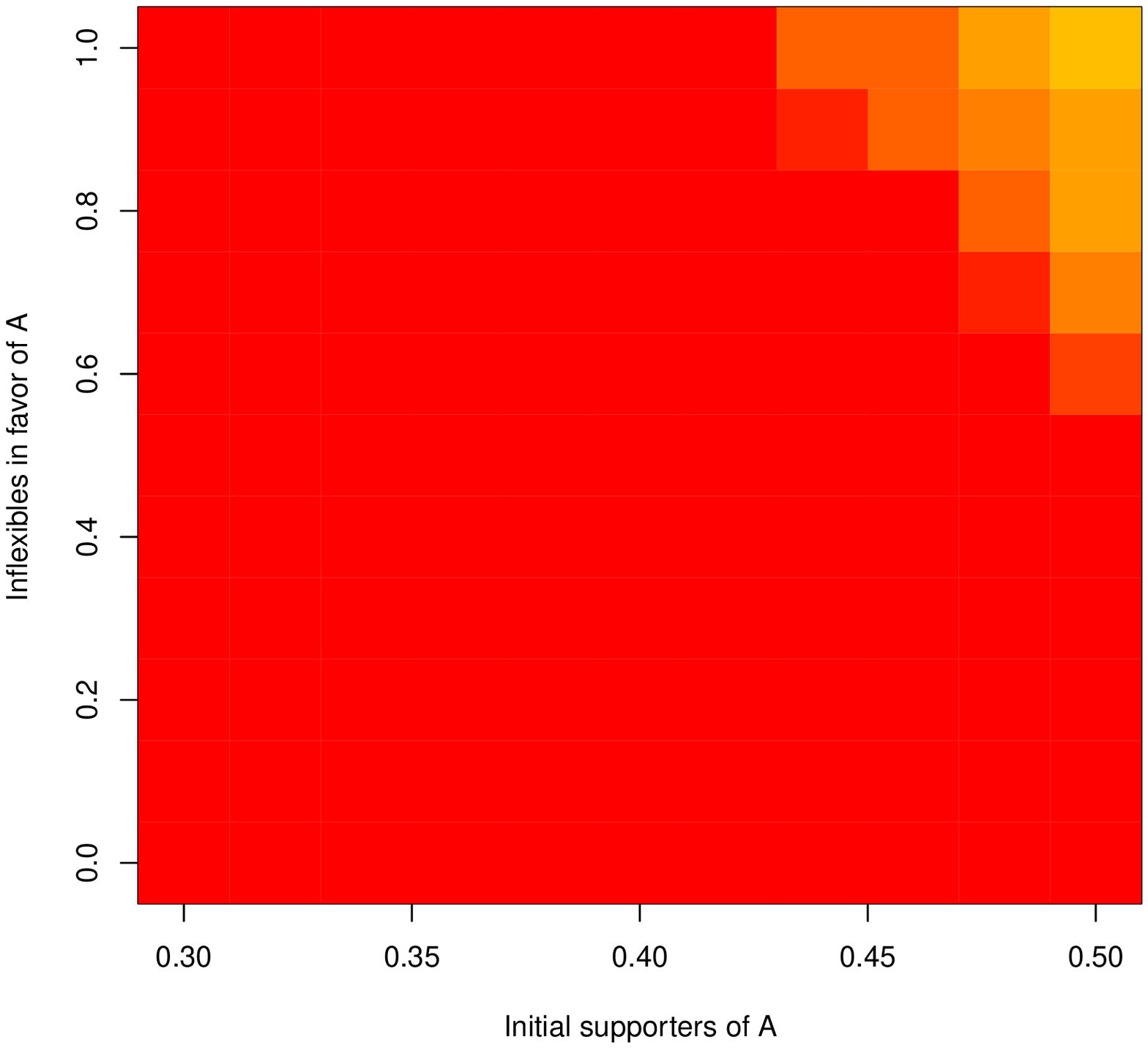,width=0.28\linewidth,clip=}
\end{tabular}
\caption{Average over 20 realizations of the final proportion of agents supporting $A$ for different values of the parameters, as a function of both $p_A$ and $I_A$, where the agent is influenced by its own opinion. Red correspond to a final state where all agents chose $B$ and white, to $A$ being chosen by all; tones of yellow show the intermediary cases. The top line figures correspond to $M=1$, the middle one to $M=5$, and the bottom one to $M=9$. Left most figures show results where $I_B=0$, central ones correspond to $I_B=50\%$, and the right ones, to $I_B=100\%$ }\label{fig:self4}
\end{figure}

The results for the case with self influence can be seen in Figure (\ref{fig:self4}). The most obvious feature is that inflexibles find it hard to convince the opposing majority, even when there are no initial inflexibles at the other side of the debate (left column). When debate is just one interaction ($M=1$), only as both $p_A$ gets close to 50\% and $I_A$ gets close to 1, we see a significant portion of population not getting completely convinced by the initial majority. As people try harder to convince each other (both for $M=5$ and $M=9$), the influence of the initial inflexibles become clearer with the appearance of a clear region where the existence of inflexibles only at the side of $A$ can make an initial majority in favor of $B$ disappear. Qualitatively, the situation is similar to the one observed for groups of size 3, altough it is required more inflexibles to have a significant effect. The other two columns show that when we have an important amount of inflexibles defending also $B$, they are capable of keeping the victory for the initial majority, as it should be expected.

\begin{figure}
\centering
\begin{tabular}{ccc}
\epsfig{file=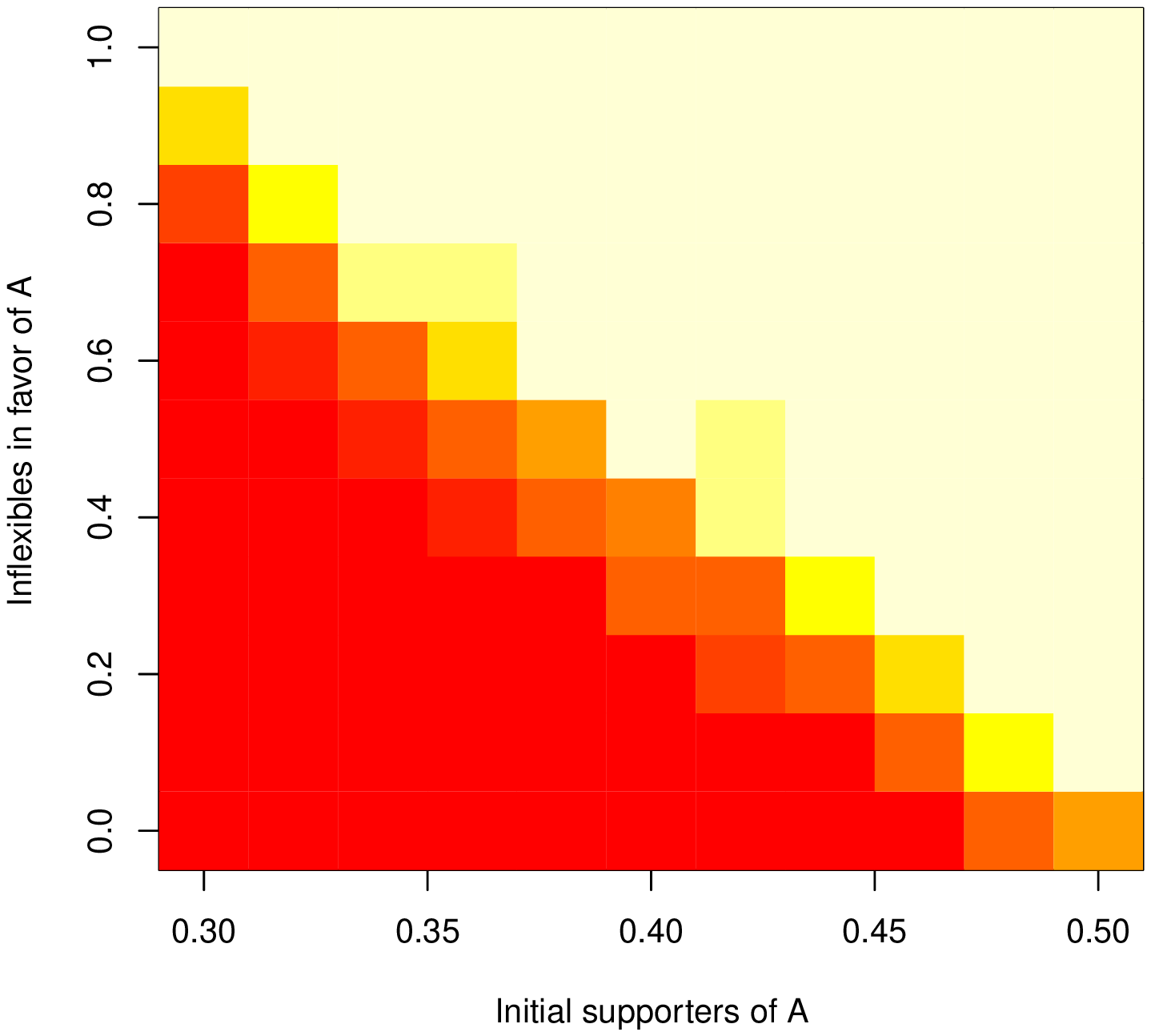,width=0.28\linewidth,clip=} & 
\epsfig{file=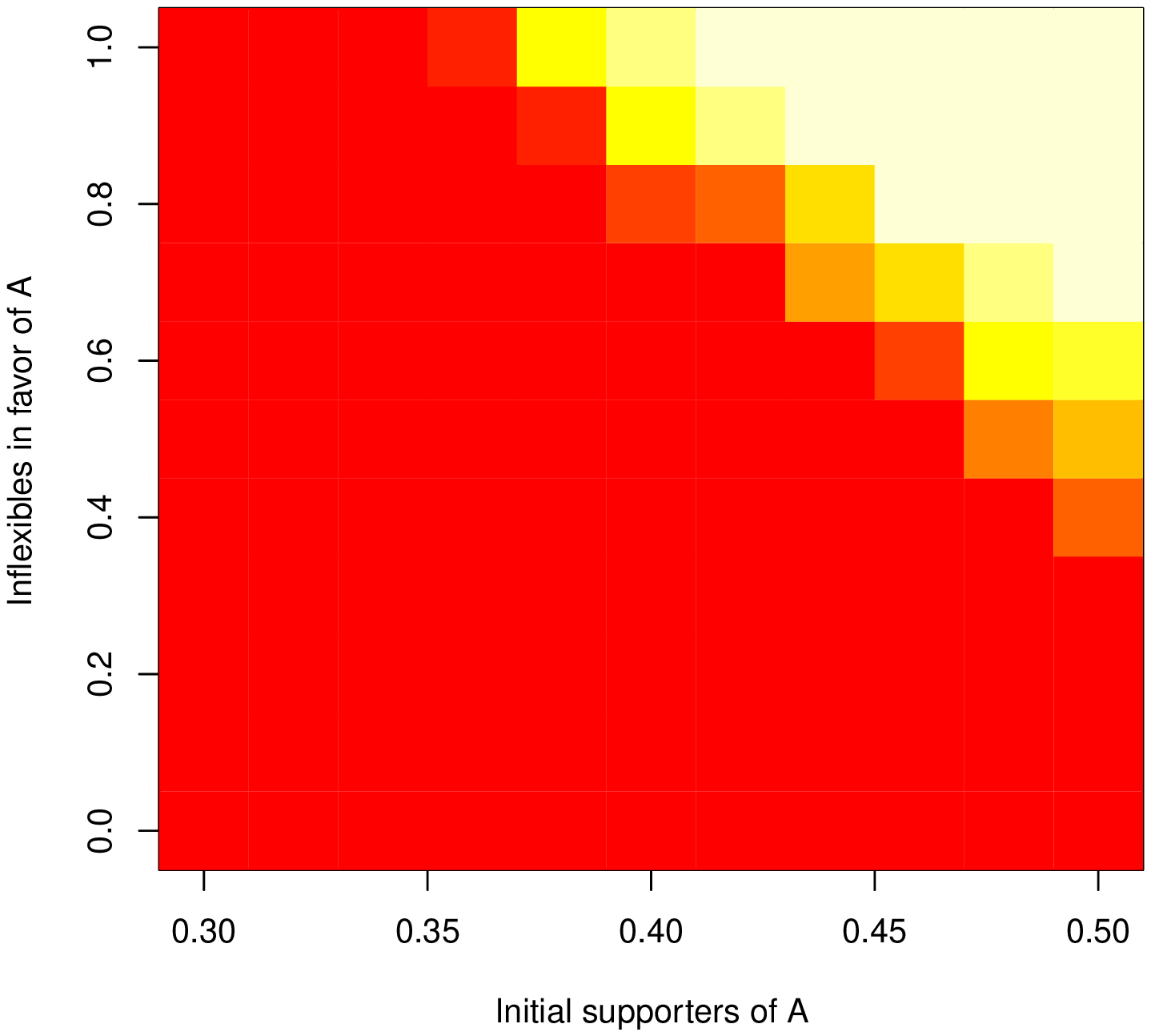,width=0.28\linewidth,clip=} & 
\epsfig{file=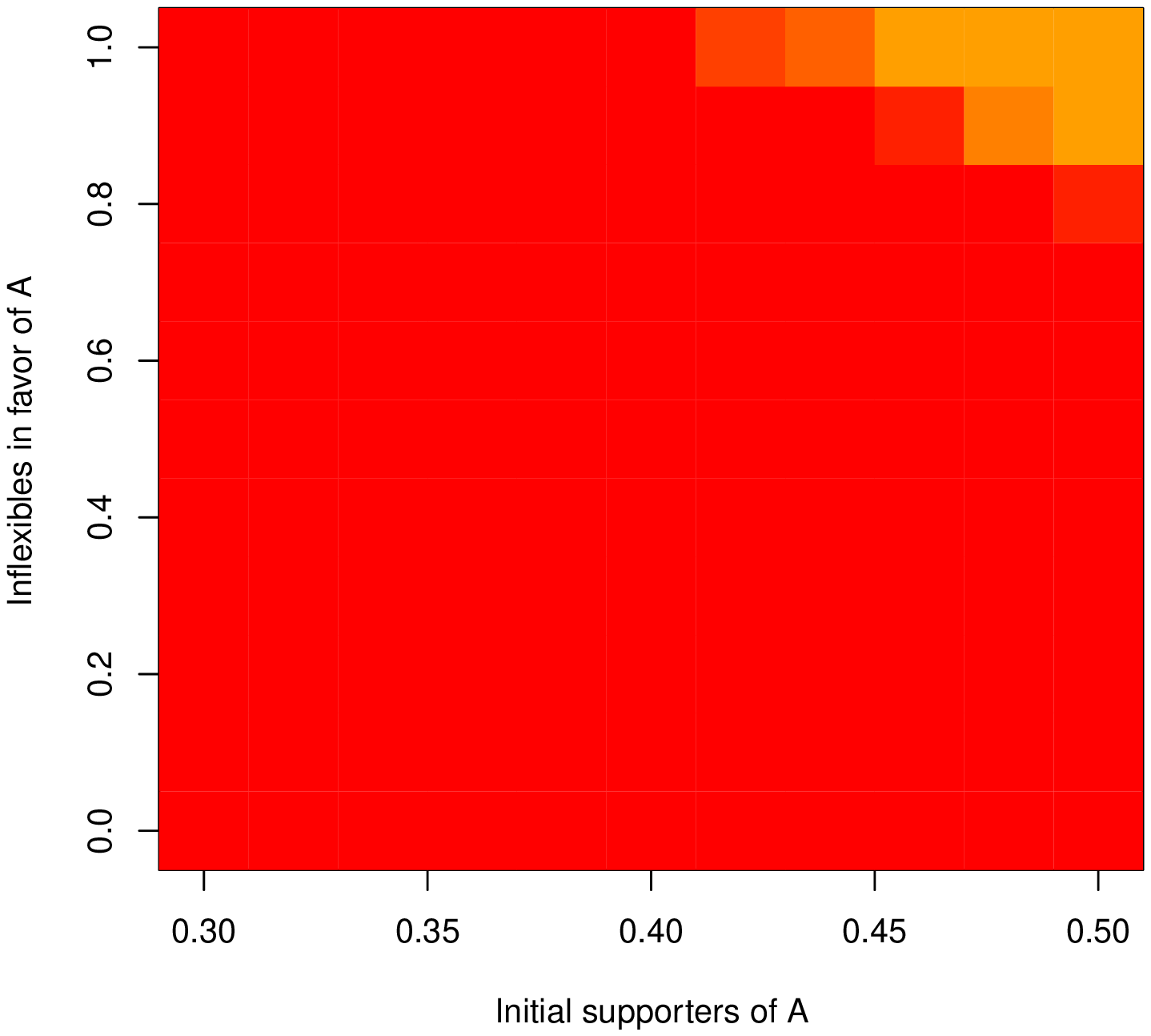,width=0.28\linewidth,clip=} \\
\epsfig{file=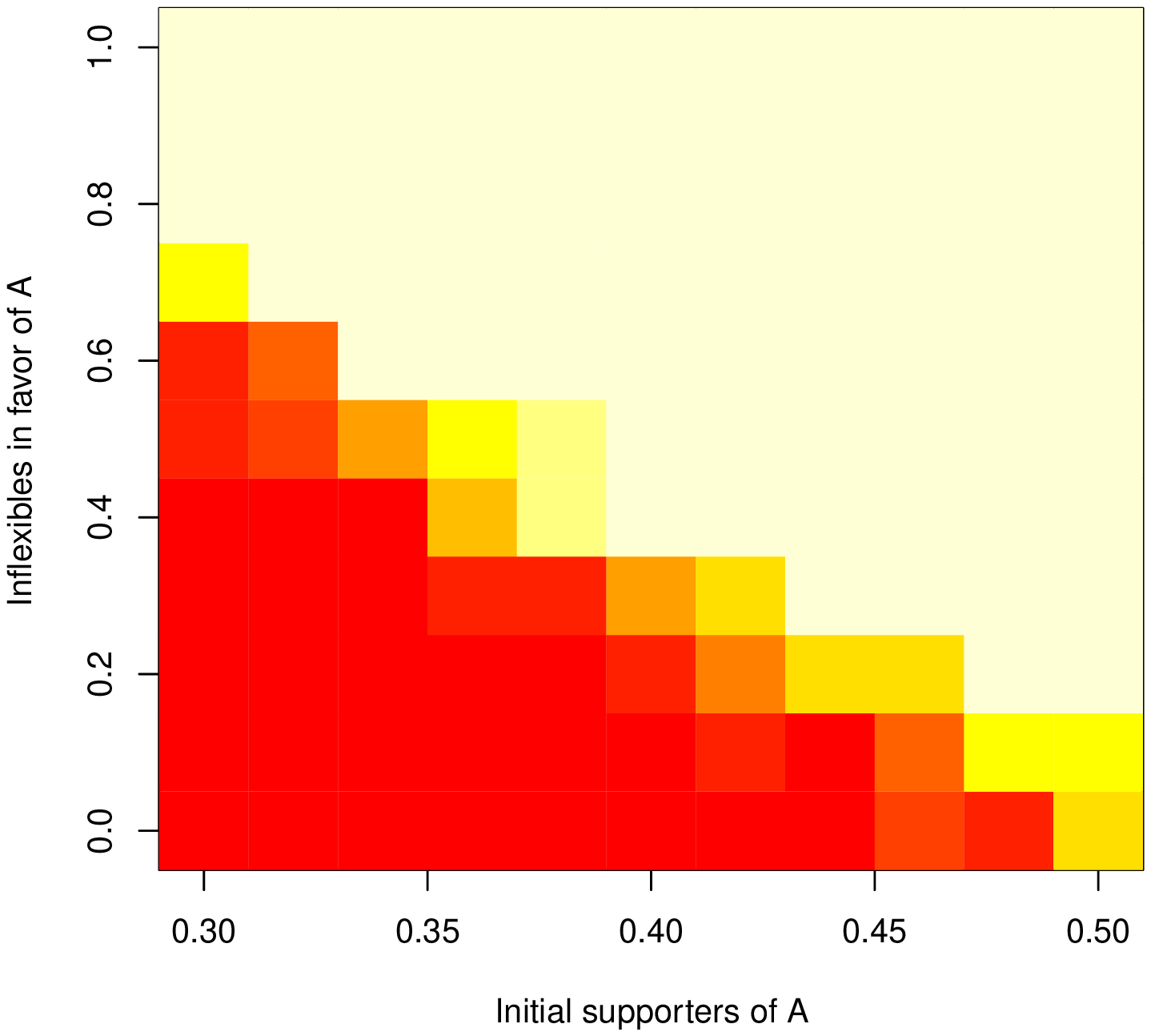,width=0.28\linewidth,clip=} & 
\epsfig{file=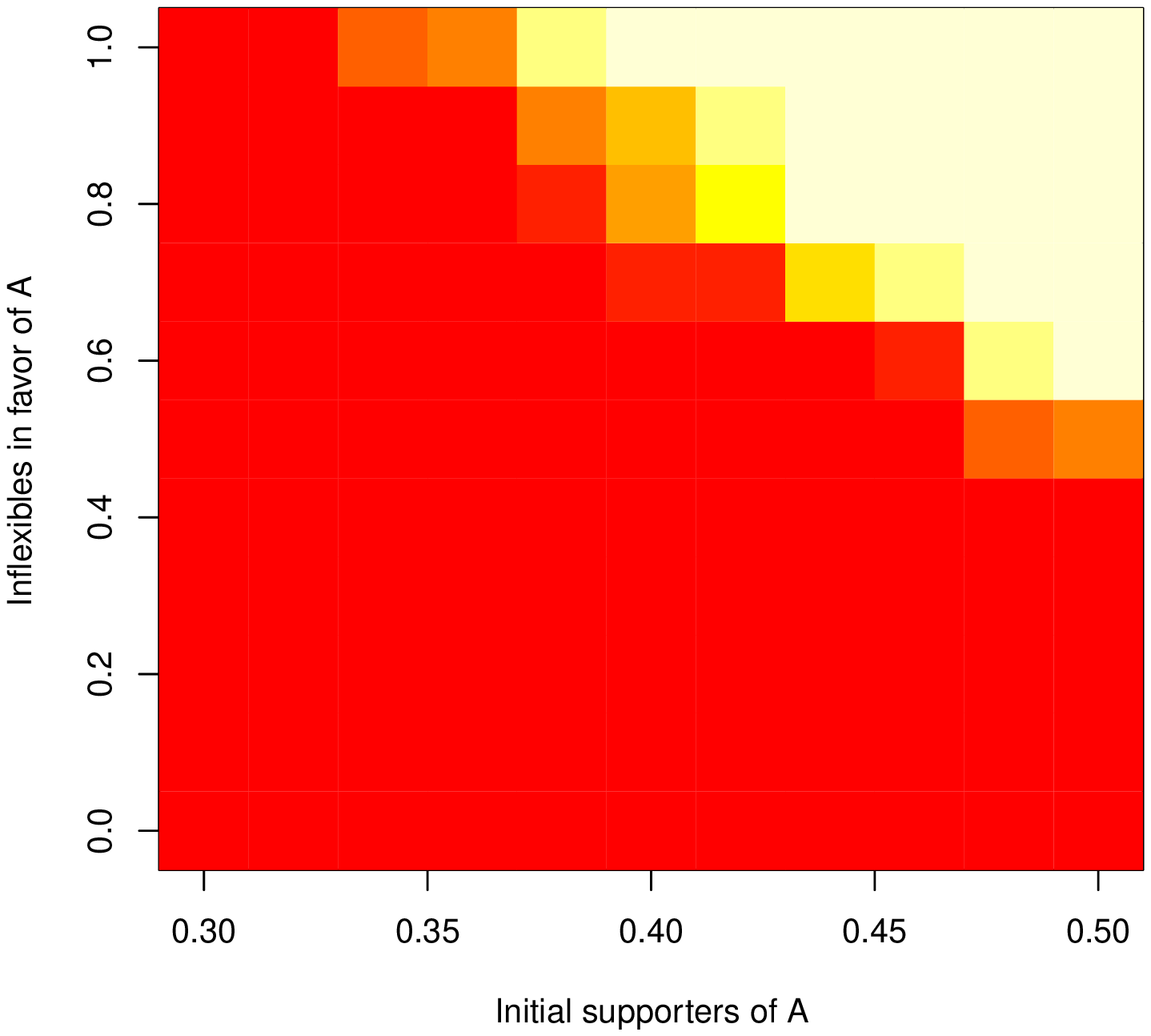,width=0.28\linewidth,clip=} & 
\epsfig{file=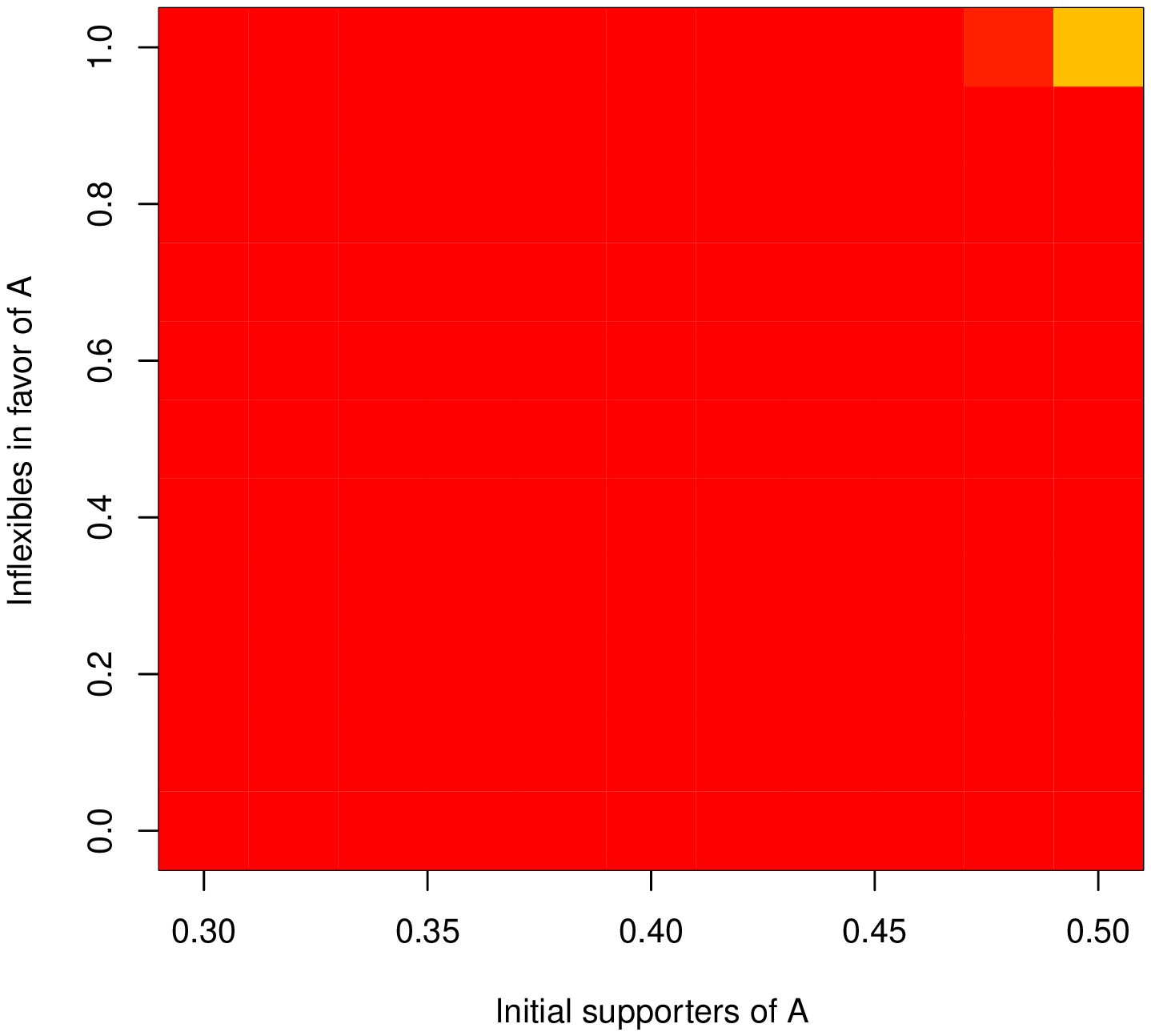,width=0.28\linewidth,clip=}\\
\epsfig{file=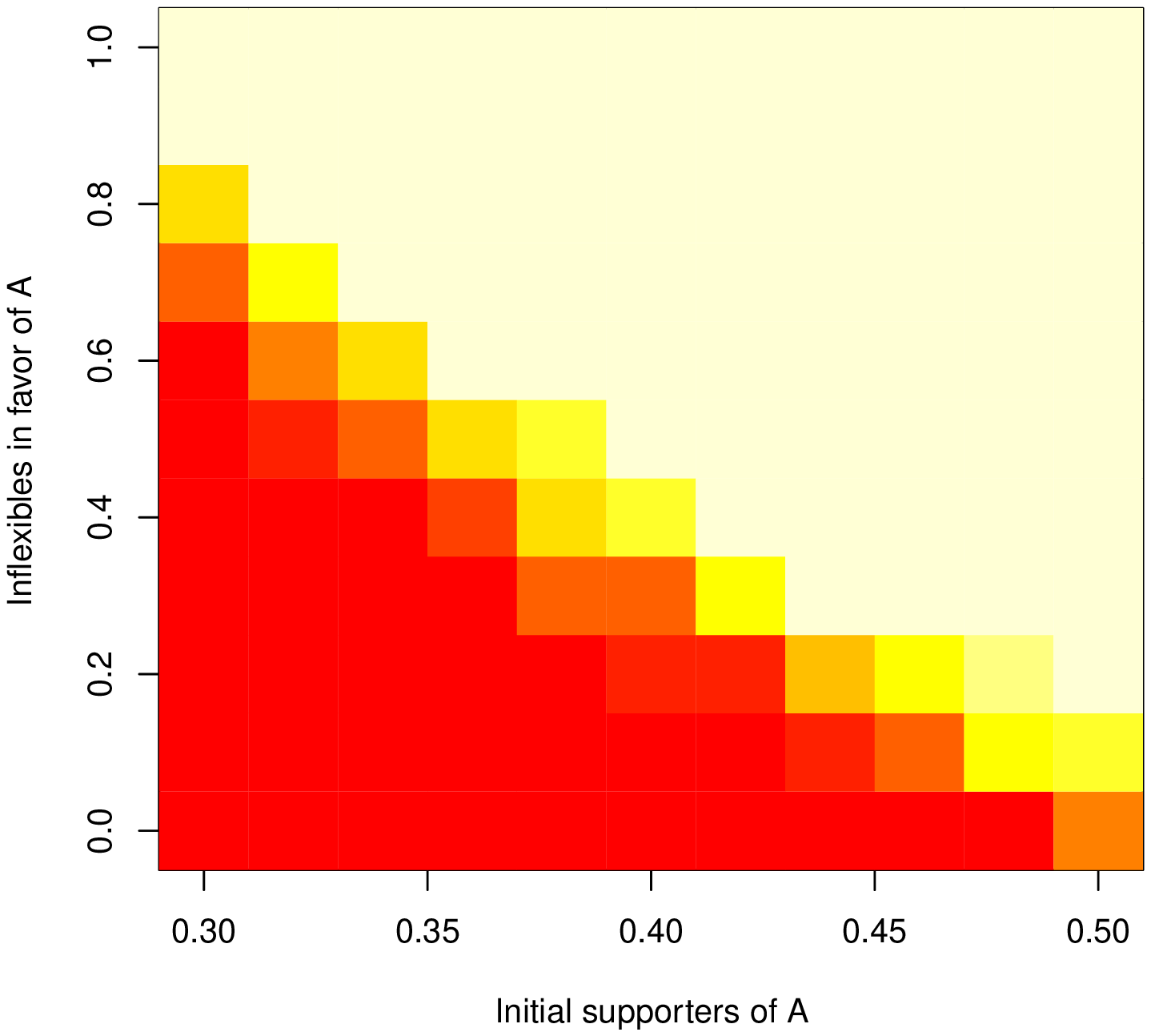,width=0.28\linewidth,clip=} & 
\epsfig{file=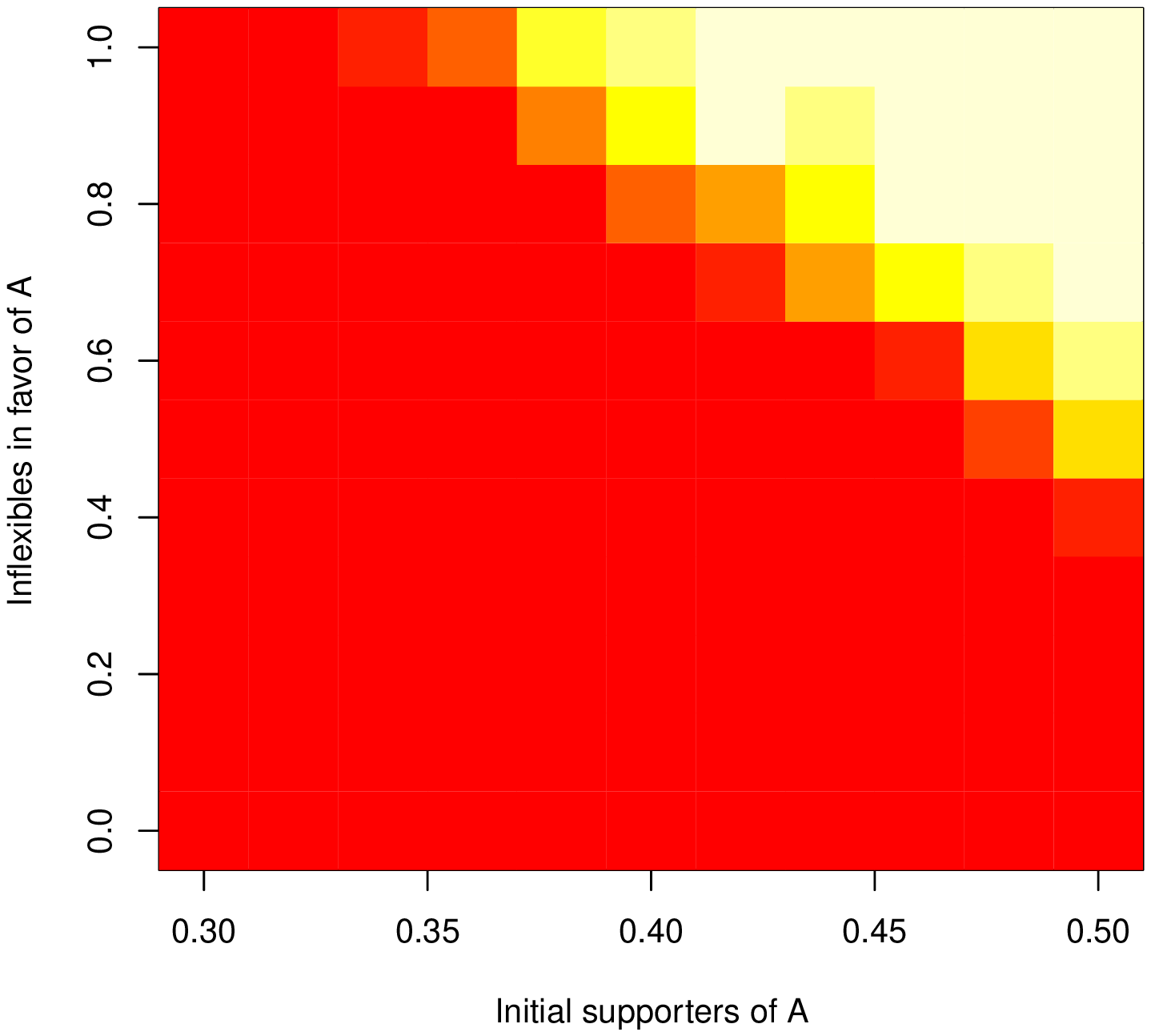,width=0.28\linewidth,clip=} & 
\epsfig{file=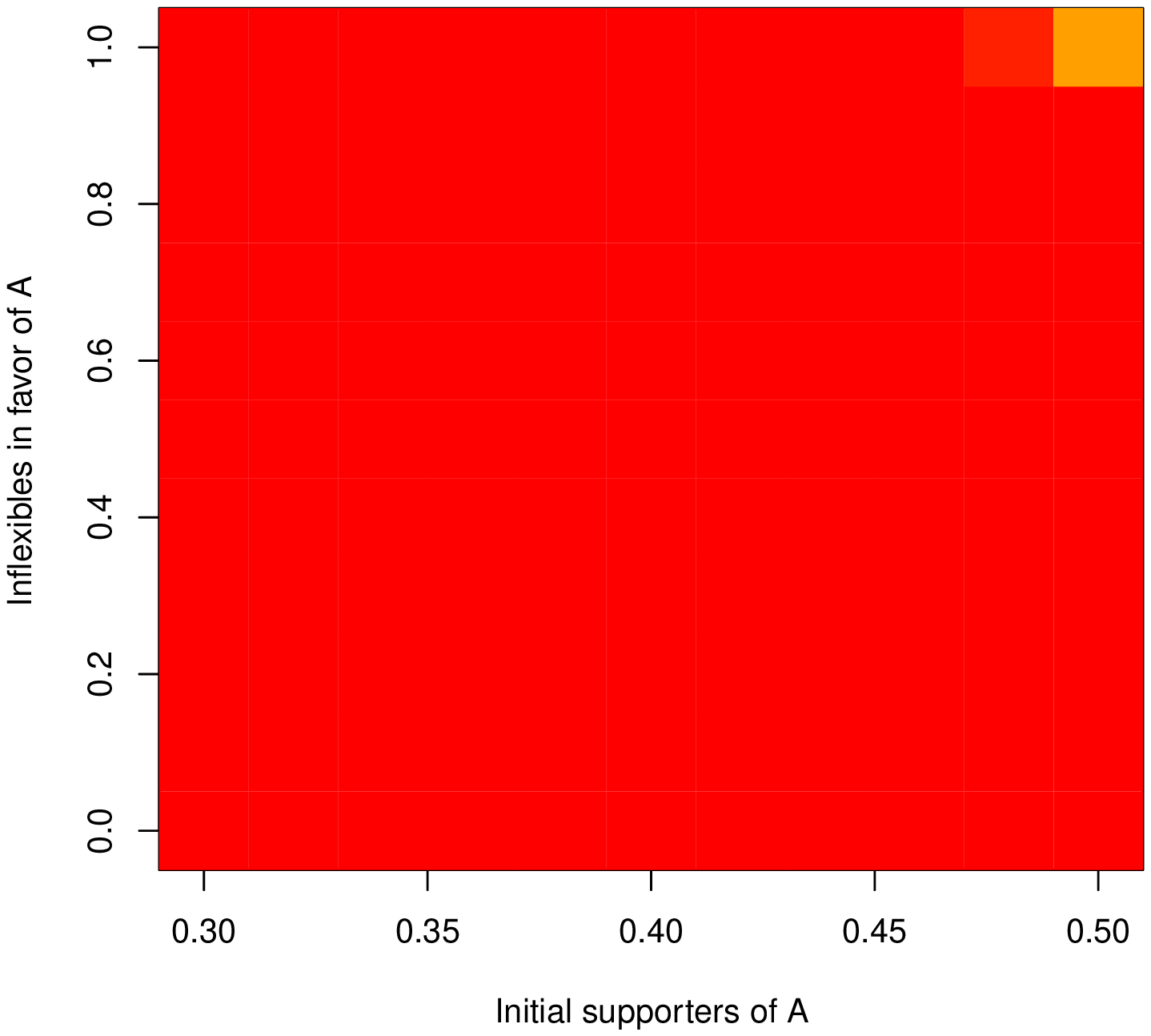,width=0.28\linewidth,clip=}
\end{tabular}
\caption{Average over20 realizations of the final proportion of agents supporting $A$ for different values of the parameters, as a function of both $p_A$ and $I_A$, where the agent is not influenced by its own opinion. Red correspond to a final state where all agents chose $B$ and white, to $A$ being chosen by all; tones of yellow show the intermediary cases. The top line figures correspond to $M=1$, the middle one, to $M=5$, and the bottom one to $M=9$. Left most figures show results where $I_B=0$, central ones correspond to $I_B=50\%$, and the right ones, to $I_B=100\%$ }\label{fig:noself4}
\end{figure}

The same analysis for the case where the agent is not influenced by its own opinion is shown in Figure (\ref{fig:noself4}). It is interesting to notice that the results are quite different from those where self influenced happened. While $M$ seems not to be far less important here, inflexibles do play a much stronger role in determining the winning side. We can see clearly the effect of increasing $I_A$ and how it allows for $A$ to win for increasingly smaller values of $p_A$. As long as inflexibles exist at only one side, their effects, in this case, are very important as for the discrete case.

\section{Conclusions}

GUF results have demonstrated that inflexibles have a major effect in determining the winning side of a public debate. However the state of indefinite inflexible could sound not realistic for most agents making those results fragile. In this work, we have shown that even a more realistic consideration of inflexibility, making it an accumulative reversible feature, preserved the major results obtained from the discrete model, at least qualitatively. Accordingly the drastic effect of inflexibility seems to a be a rather robust ingredient of social debates. Here, we were able to estimate the effect of inflexible tuning the local debate length via the instrumental parameter $M$, which reinforces or weakens the agent respective convictions. Larger $M$ provides a tiny inflexible minority with the power to convince a majority to shift opinion and get aligned along the minority view.

More investigation could be of interest to determine more clearly the fact that an agent influencing itself sounds as a good psychological description although  it appears illogical with respect the position of listening other arguments to make up one's won choice.

\section{Acknowledgements.} One of the authors (ACRM) would like to thank the Funda\c{c}\~ao de Amparo a Pesquisa do Estado de S\~ao Paulo (FAPESP) for the support to the work under grant 2009/08186-0.

\bibliographystyle{unsrt}
\bibliography{biblio}

\end{document}